\documentclass[pdflatex,sn-mathphys-num]{sn-jnl}
\usepackage{graphicx}%
\usepackage{subcaption}%

\usepackage{multirow}%
\usepackage{amsmath,amssymb,amsfonts}%
\usepackage{amsthm}%
\usepackage{mathrsfs}%
\usepackage[title]{appendix}%
\usepackage{xcolor}%
\usepackage{textcomp}%
\usepackage{manyfoot}%
\usepackage{booktabs}%
\usepackage{algorithm}%
\usepackage{float}
\usepackage{placeins}
\usepackage{algorithmicx}%
\usepackage{algpseudocode}%
\usepackage{listings}%
\theoremstyle{thmstyleone}%
\newtheorem{theorem}{Theorem}%
\newtheorem{corollary}[theorem]{Corollary}
\theoremstyle{thmstyletwo}%
\newtheorem{remark}{Remark}%

\theoremstyle{thmstylethree}%
\newtheorem{theoremA}{Theorem}

\begin{document}

\title[Article Title]{Parsimonious Modeling of Periodic Time Series Using Fourier and Wavelet Techniques }
\author*[1]{\fnm{Rhea} \sur{Davis}}\email{rheadavisc@gmail.com}

\author[2]{\fnm{N.} \sur{Balakrishna}}\email{nb@cusat.ac.in, bala@iittp.ac.in}

\affil*[1]{\orgdiv{Department of Statistics}, \orgname{Cochin University of Science and Technology}, \orgaddress{\city{Kochi},  \state{Kerala}, \country{India}}}

\affil[2]{\orgdiv{Department of Mathematics and Statistics}, \orgname{Indian Institute of Technology}, \orgaddress{\city{Tirupati}, \state{Andhra Pradesh}, \country{India}}}

\abstract{This paper proposes Fourier-based and wavelet-based techniques for analyzing periodic financial time series. Conventional models such as the periodic autoregressive conditional heteroscedastic (PGARCH) and periodic autoregressive conditional duration (PACD) often involve many parameters. The methods put forward here resulted in more parsimonious models with increased forecast efficiency. The effectiveness of these approaches is demonstrated through simulation and data analysis studies.}


\maketitle

\section{Introduction}
Many economic and financial time series exhibit seasonal patterns. For example, stock price movements show the presence of calendar effects, which are expected to have patterns related to the month, day, or hour. For instance, in \citeauthor{floros2008monthly} (\citeyear{floros2008monthly}) it is mentioned that the returns in the first half of a month are significantly higher than those in the second half. This observation has also been reported in \citeauthor{rosenberg2004monthly} (\citeyear{rosenberg2004monthly}). Some of the other references that studied the monthly effect are  \citeauthor{rozeff1976capital} (\citeyear{rozeff1976capital}) and \citeauthor{gallant1992stock} (\citeyear{gallant1992stock}). For daily returns data, it is found in \citeauthor{chiah2019day} (\citeyear{chiah2019day}) that there is a strong preference for speculative stocks on Friday and non-speculative stocks on Monday. Similarly, \citeauthor{chiah2021tuesday} (\citeyear{chiah2021tuesday}) observes that the `Tuesday Blues' phenomenon is predominantly present among speculative Australian stocks. The day-of-the-week effect in financial series has also been investigated by \citeauthor{ul2021mood} (\citeyear{ul2021mood}), \citeauthor{chan2012day} (\citeyear{chan2012day}), and \citeauthor{romero2010intraday} (\citeyear{romero2010intraday}). Upon analyzing the intraday return data, \citeauthor{pigorsch2024reversal} (\citeyear{pigorsch2024reversal}) realized that it is the returns on Monday afternoon and not the morning that reverse over the remainder of the week. \citeauthor{ma2025night} (\citeyear{ma2025night}) studied the impact of night trading and concluded that after the introduction of night trading, the returns of the night's first half hour have considerable predictive power, while the returns of the day's first half hour were a significant predictor before the commencement of night trading. Some of the widely cited works on intraday effect are \citeauthor{wood1985investigation} (\citeyear{wood1985investigation}) and \citeauthor{french1986stock} (\citeyear{french1986stock}). 


A major share of literature on the financial time series focuses on modeling the stochastic volatility in terms of the return series $\{y_t\}$, which includes different variants of generalized autoregressive conditional heteroscedastic $(p, q)$ (GARCH$(p, q)$) model (see \citeauthor{bollerslev1986generalized}, \citeyear{bollerslev1986generalized}) defined by
\begin{equation}
    \begin{split}
        y_t &= \epsilon_t \sqrt{h_t}\\
        \text{and} \quad h_t &= \omega+\sum\limits_{i=1}^p \alpha_iy_{t-i}^2+\sum\limits_{j=1}^q \beta_jh_{t-j}, \, t= 0, 1, \dotsc, 
    \end{split}
\end{equation}
where $\{\epsilon_t\}$ is a sequence of real-valued random variables with $E(\epsilon_t) = 0$ and $Var(\epsilon_t) = 1$. It is assumed that $\omega>0, \alpha_i \geq 0, i = 1,\dotsc,p,$ and $\beta_j \geq 0, j = 1,\dotsc,q$, to ensure the positivity of $h_t$. An equally important model is the autoregressive conditional duration $(r,s)$ (ACD $(r, s)$) model (see \citeauthor{engle1998autoregressive}, \citeyear{engle1998autoregressive}) for modeling the durations $\{u_t\}$ between the transactions whose structure is described by 
\begin{equation}
    \begin{split}
        u_t &= \psi_t \xi_t\\
        \text{and} \quad \psi_t &= \lambda+\sum\limits_{i = 1}^s \gamma_iu_{t-i}+\sum\limits_{j=1}^r \delta_j \psi_{t-j}, \, t = 0,1,\dotsc,
    \end{split}
\end{equation}
where $\{\xi_t\}$ is a sequence of non-negative random variables with mean 1 and variance $\sigma^2$. The positivity of $\psi_t$ is ensured by assuming $\lambda>0, \gamma_i \geq 0, i = 1, \dotsc, s,$ and $\delta_j \geq 0, j = 1, \dotsc, r$. To illustrate our proposed approach, we focus on the setting where $p=q=1$ and $r=s=1$, corresponding to the simplest case of both processes.

The pervasive presence of periodic patterns in financial time series has stimulated many researchers to incorporate periodicity into the various characteristics of existing models. \citeauthor{bollerslev1996periodic} (\citeyear{bollerslev1996periodic}) formulated the periodic generalized autoregressive conditional heteroscedastic (PGARCH) model to capture the periodicity in volatility by including periodically varying parameters of period $\nu$ in the conditional-variance equation. The $\text{PGARCH}_{\nu}(1,1)$ model is (\citeauthor{bollerslev1996periodic}, \citeyear{bollerslev1996periodic})
  \begin{equation}\label{PGARCHeq}
    \begin{split}
        y_t &= \epsilon_t \sqrt{h_t}\\
        \text{and} \quad h_t &= \omega_t+\alpha_ty_{t-1}^2+\beta_t h_{t-1}, \, t = 0, 1, \dotsc, 
      \end{split}
    \end{equation}
    where (i) $\omega_t = \omega_{t+k\nu}, k \in \mathbb{Z},$ (ii) $\alpha_t = \alpha_{t+k\nu}, k \in \mathbb{Z},$ and (iii) $\beta_t = \beta_{t+k\nu}, k \in \mathbb{Z}$. Let $\boldsymbol{\omega} = (\omega_0,\dotsc,\omega_{\nu-1})^T, \boldsymbol{\alpha} = (\alpha_0,\dotsc,\alpha_{\nu-1})^T$, and $\boldsymbol{\beta} = (\beta_0,\dotsc,\beta_{\nu-1})^T$, where $T$ denotes the transpose. Also, $\{\epsilon_t\}$ is an independent and identically distributed sequence of real-valued random variables with $E(\epsilon_t) = 0$, $Var(\epsilon_t) = 1$, and $\epsilon_k$ is independent of $y_t$, for all $k>t$. Let $y_0,\dotsc,y_{N\nu-1}$ be the data generated from \eqref{PGARCHeq}, where $N$ is the number of cycles. It is assumed that $\omega_t >0, \alpha_t \geq 0$, and $\beta_t \geq 0$ to ensure positivity of $h_t$. Note that PGARCH reduces to the widely used GARCH model when $\nu = 1$. 
    
    The well-established autoregressive conditional duration (ACD) model to analyze positive-valued time series was modified by \citeauthor{aknouche2022periodic} (\citeyear{aknouche2022periodic}) through the inclusion of periodically varying parameters of period $\nu$ in the conditional duration equation, resulting in the periodic autoregressive conditional duration (PACD) model. The $\text{PACD}_{\nu}(1,1)$ model is described by (\citeauthor{aknouche2022periodic}, \citeyear{aknouche2022periodic})
          \begin{equation}\label{PACD eq}
          \begin{split}
              u_t &= \psi_t \xi_t\\
          \text{and} \quad  
              \psi_t &= \lambda_t + \gamma_t u_{t-1} + \delta_t \psi_{t-1}, \, t = 0, 1, \dotsc,
              \end{split}
          \end{equation}
          where (i) $\lambda_t = \lambda_{t+k\nu}, k \in \mathbb{Z} $, (ii) $\gamma_t = \gamma_{t+k\nu}, k \in \mathbb{Z}$, and (iii) $\delta_t = \delta_{t+k\nu}, k \in \mathbb{Z} $. Let $\boldsymbol{\lambda} = (\lambda_0,\dotsc,\lambda_{\nu-1})^T, \boldsymbol{\gamma} = (\gamma_0,\dotsc,\gamma_{\nu-1})^T$, and $\boldsymbol{\delta} = (\delta_0,\dotsc,\delta_{\nu-1})^T$. $\{\xi_t\}$ is a sequence of non-negative random variables with mean 1 and variance
        $\sigma_t^2$ satisfying $\sigma_t^2 = \sigma_{t+k\nu}^2, k \in \mathbb{Z}$. Define $\boldsymbol{\sigma}^2 = (\sigma_0^2,\dotsc,\sigma_{\nu-1}^2)$. Let $u_0,\dotsc,u_{N\nu-1}$ be the data generated from \eqref{PACD eq}.
        The positivity of $\psi_t$ is ensured by assuming $\lambda_t >0, \gamma_t \geq 0$, and $\delta_t \geq 0$. The standard ACD model is obtained when $\nu = 1$. A key limitation of the PGARCH as well as PACD models is the presence of an extensive set of parameters, which can lead to suboptimal forecasting performance. For example, in contexts where day-of-the-week patterns ($\nu = 7$) are observed, the PGARCH model includes 21 parameters ($7 \times 3$), while the PACD model comprises 28 parameters ($7 \times 4$). Hence, it is crucial to design techniques to reduce the number of parameters while still capturing the underlying dependency structure and enhancing forecasting performance. 
        
        For models \eqref{PGARCHeq} and \eqref{PACD eq}, we do not insist on the exact form of the density for either $\{\epsilon_t\}$ or $\{\xi_t\}$. However, parameter estimation for the PGARCH and PACD models is carried out using the quasi-maximum likelihood estimation method. It is customary to develop quasi-maximum likelihood estimator (QMLE) by assuming a standard normal distribution for real-valued innovations as suggested in \citeauthor{lee1994asymptotic} (\citeyear{lee1994asymptotic}). Similarly, when the innovations are positive, as in the PACD case, a unit exponential distribution is commonly assumed (see \citeauthor{pacurar2008autoregressive}, \citeyear{pacurar2008autoregressive}). The Fourier and wavelet methods proposed for reducing the number of parameters require identifying significant coefficients. This identification is performed through hypothesis testing based on the asymptotic distributions of the Fourier and wavelet coefficients, which are derived from the asymptotic normality of the QMLEs. Let $\boldsymbol{\theta} = (\theta_0, \theta_1, \dotsc, \theta_{3\nu-1})^T= (\omega_0, \alpha_0, \beta_0,\dotsc, \omega_{\nu-1}, \alpha_{\nu-1}, \beta_{\nu-1})^T$. If $\hat{\boldsymbol{\theta}}$ is the QMLE of $\boldsymbol{\theta}$, then $\hat{\boldsymbol{\theta}}$ is asymptotically normal with the asymptotic variance-covariance matrix $\Sigma$, as stated in Theorem \ref{PGARCHasynormthm}. Since $\Sigma$ is a block diagonal matrix, the techniques employed to reduce the number of parameters can be applied separately to the vectors $\boldsymbol{\omega}, \boldsymbol{\alpha}$, and $\boldsymbol{\beta}$, whose corresponding asymptotic variance-covariance matrices are $\Sigma_{\boldsymbol{\omega}}, \Sigma_{\boldsymbol{\alpha}}$, and $\Sigma_{\boldsymbol{\beta}}$, respectively, as described in Corollary \ref{PGARCHestasynormthm}. Analogous results hold for the vectors $\boldsymbol{\lambda}, \boldsymbol{\gamma}$, and $\boldsymbol{\delta}$ in the PACD model. The relevant theoretical results used in the paper are listed in the Appendix.

The structure of the article is as follows. In Section 2, we introduce the proposed
Fourier-based techniques, while Section 3 introduces the wavelet-based framework developed in this study. Simulation studies of the proposed methods are presented in Section 4. Section 5 illustrates the application of these techniques to real-world data. Finally, Section 6 concludes the study. Additional supporting material is provided in the Appendix.

\section{Fourier Analysis of Parameter Vectors}
Since the parameter vectors in both PGARCH and PACD models exhibit periodicity, we first consider a Fourier-based method as it is a natural tool for analyzing such vectors. In the context of periodic vectors, representing them using a Fourier series is particularly appropriate, as the coefficients associated with the sine and cosine terms indicate the strength of the corresponding frequency components. Often, a reasonable approximation can be achieved by retaining only a subset of these components. The objective, therefore, is to represent a parameter vector using only its most significant frequency components.

Let $\mathbf{X} = (X_0,\dotsc, X_{\nu-1})^T$ denote any one of the parameter vectors $\boldsymbol{\omega}, \boldsymbol{\alpha},\boldsymbol{\beta},\boldsymbol{\lambda},\boldsymbol{\gamma},\boldsymbol{\delta}$, and $\boldsymbol{\sigma}^2$. The Fourier representation of $\mathbf{X}$ is (\citeauthor{anderson2021parsimonious}, \citeyear{anderson2021parsimonious})
\begin{equation} \label{DFT}
    X_t = c_{\textbf{X},0}+\sum_{r=1}^{\ell}\left\{c_{\textbf{X},r}\cos\left(\frac{2\pi rt}{\nu}\right)+s_{\textbf{X},r}\sin\left(\frac{2\pi r t}{\nu}\right)\right\}, \, 0 \leq t \leq \nu-1,
\end{equation}
where $c_{\textbf{X},r}$ and $s_{\textbf{X},r}$ are Fourier coefficients, and $\ell$ is $\nu/2$, for $\nu$ even and $(\nu -1)/2$, for $\nu$ odd. 
Accordingly, the Fourier coefficient vector is defined as (\citeauthor{anderson2021parsimonious}, \citeyear{anderson2021parsimonious})
\begin{equation}\label{Fouriervec}
\textbf{f}_{\textbf{X}} = \begin{cases}
    (c_{\textbf{X},0},c_{\textbf{X},1},s_{\textbf{X},1},\dotsc,c_{\textbf{X},(\nu-1)/2},s_{\textbf{X},(\nu-1)/2})^{T} \quad (\nu \, \text{odd}) \\
    (c_{\textbf{X},0},c_{\textbf{X},1},s_{\textbf{X},1},\dotsc,s_{\textbf{X},(\nu /2 -1)},c_{\textbf{X},(\nu /2)})^{T} \quad \quad (\nu \, \text{even})
\end{cases}.
\end{equation}
Let $\mathcal{L}$ and $\mathcal{U}$ be the matrices (\citeauthor{anderson2021parsimonious}, \citeyear{anderson2021parsimonious})
\begin{equation}\label{A.L.exp}
    \mathcal{L} = \begin{cases}
        \text{diag} (\nu^{-1/2}, \sqrt{2/ \nu}, \dotsc, \sqrt{2/ \nu} ) \quad \quad \quad \quad (\nu \, \text{odd}) \\
        \text{diag} (\nu^{-1/2}, \sqrt{2/ \nu}, \dotsc, \sqrt{2/ \nu}, \nu^{-1/2} ) \quad (\nu \, \text{even})
    \end{cases}, 
\end{equation}
and 
\begin{equation}\label{A.U.exp}
    \mathcal{U} = \nu^{-1/2}\left(e^{\frac{-i2\pi rt}{\nu}}\right),
\end{equation}
respectively. Additionally, consider the matrix $\mathcal{P}$ whose $(\ell,j)^{th}$ element is defined by (\citeauthor{anderson2021parsimonious}, \citeyear{anderson2021parsimonious})
\begin{equation}\label{A.P.exp}
    [\mathcal{P}]_{\ell j}= \begin{cases}1 & \text { if } \ell=j=0 ; \\ 2^{-1 / 2} & \text { if } \ell=2 r-1 \text { and } j=r \text { for some } 1 \leq r \leq[(v-1) / 2]; \\ 2^{-1 / 2} & \text { if } \ell=2 r-1 \text { and } j=v-r \text { for some } 1 \leq r \leq[(v-1) / 2]; \\ i 2^{-1 / 2} & \text { if } \ell=2 r \text { and } j=r \text { for some } 1 \leq r \leq[(v-1) / 2] ; \\ -i 2^{-1 / 2} & \text { if } \ell=2 r \text { and } j=v-r \text { for some } 1 \leq r \leq[(v-1) / 2] ; \\ 1 & \text { if } \ell=v-1 \text { and } j=v / 2 \text { and } v \text { is even; and } \\ 0 & \text { otherwise. }\end{cases}.
\end{equation}
Then, \eqref{Fouriervec} has the matrix representation (\citeauthor{anderson2021parsimonious}, \citeyear{anderson2021parsimonious}),
\begin{equation}\label{fx exp}
\textbf{f}_{\textbf{X}} = \mathcal{L}\mathcal{P}\mathcal{U}\mathbf{X}. 
\end{equation}
The vector $\textbf{X}$ can be reconstructed from $\textbf{f}_{\textbf{X}}$ by, 
\begin{equation*}
    \textbf{X} = \mathcal{U}^{\dag}\mathcal{P}^{\dag}\mathcal{L}^{-1}\textbf{f}_{\textbf{X}},
\end{equation*}
where $\mathcal{A}^{\dag}$ denotes the conjugate transpose of unitary matrix $\mathcal{A}$. Thus, $\textbf{X}$ and $\textbf{f}_{\textbf{X}}$ provide different aspects of the same object. 

The periodic autoregressive moving average model (PARMA), similar to the PGARCH and PACD models, is characterized by many parameters. However, since these parameters typically exhibit smooth periodic variation, previous studies \citeauthor{anderson1993asymptotic} (\citeyear{anderson1993asymptotic}), \citeauthor{vecchia1985maximum} (\citeyear{vecchia1985maximum}), \citeauthor{anderson2007fourier} (\citeyear{anderson2007fourier}), and \citeauthor{tesfaye2011asymptotic} (\citeyear{tesfaye2011asymptotic}) have successfully approximated them using a limited number of significant Fourier coefficients. Inspired by these results, we propose applying the Fourier-based technique to reduce the number of parameters in the PGARCH and PACD models.
 Suppose the estimator $\hat{\textbf{X}}$ of $\mathbf{X}$ is asymptotically normal as described in 
\begin{equation}\label{Xasynorm}
    N^{1/2}(\hat{\textbf{X}}-\textbf{X}) \xrightarrow{d} \mathcal{N}(\mathbf{0},\Gamma_{\mathbf{X}}).
\end{equation}
Then, according to Theorem \ref{A.broc.asymnormaltheorem}, the associated vector of Fourier coefficients $\hat{\textbf{f}}_{\textbf{X}}$ is also asymptotically normally distributed as shown in \citeauthor{anderson2021parsimonious} (\citeyear{anderson2021parsimonious})
\begin{equation}\label{Fourierasynorm}
N^{1/2}(\hat{\textbf{f}}_{\textbf{X}} - \mathbf{f}_{\mathbf{X}}) \xrightarrow{d} \mathcal{N}(\mathbf{0}, R_{\mathbf{f}_{\mathbf{X}}}),
\end{equation}
with $R_{\mathbf{f}_{\mathbf{X}}} = \mathcal{L}\mathcal{P}\mathcal{U} \Gamma_{\mathbf{X}} \mathcal{U}^{\dag} \mathcal{P}^{\dag} \mathcal{L}$. Our objective is to represent each parameter vector in a PGARCH or PACD model using only a small number of statistically significant Fourier coefficients. Significant coefficients are identified via a hypothesis testing procedure based on the asymptotic distribution given in \eqref{Fourierasynorm}. The null hypothesis of the test is $H_0: \mathbf{f}_{\mathbf{X}} = (c_{\mathbf{X},0},0,\dotsc,0)^T$. Under this hypothesis, we have $X_i = c_{\mathbf{X},0}, 0\leq i \leq \nu-1$. Suppose all parameter vectors are reduced to constants. In that case, this implies that the parameters are time-invariant, thereby reducing the PGARCH or PACD model to the standard GARCH or ACD model, respectively. Now, \eqref{Fourierasynorm} yields 
\begin{equation}
    N^{1/2}(\hat{{f}}_{\textbf{X},i}-f_{\mathbf{X},i}) \xrightarrow{d} \mathcal{N}(0,[R_{\mathbf{f}_{\mathbf{X}}}]_{ii}), \quad 0 \leq i \leq \nu-1,
\end{equation}
where $f_{\textbf{X},i}$ is the $i^{th}$ element of $\textbf{f}_{\textbf{X}}$, and $[R_{\textbf{f}_{\textbf{X}}}]_{ii}$ denotes the $i^{th}$ diagonal element of matrix $R_{\textbf{f}_{\textbf{X}}}$. In this context, the hypothesis test is formulated with null and alternative hypotheses as $H_0:|f_{\textbf{X},i}|=0$ and $H_1:|f_{\textbf{X},i}|\neq0$, for $i = 1, 2, \dotsc, \nu-1$, respectively. The test statistic is defined by 
\begin{equation}
    Z_{\mathbf{f},i} = \frac{\hat{f}_{\mathbf{X},i}}{\sqrt{[R_{\mathbf{f}_{\mathbf{X}}}]_{ii}/N}}, \quad 1 \leq i \leq \nu-1,
\end{equation} 
and let $\Phi$ denote the cumulative distribution function of the standard normal distribution. Since multiple hypotheses are being tested simultaneously, the Bonferroni correction (see \citeauthor{howell2012statistical}, \citeyear{howell2012statistical}) controls the family-wise error rate. Hence, when $|Z_{\mathbf{f},i}|>\Phi^{-1}(\alpha/(2(\nu-1)))$, the null hypothesis is rejected, and the corresponding Fourier coefficient is considered significant and retained. Otherwise, it is set to zero. We expect only a small number of coefficients to satisfy the significance criterion, resulting in a parsimonious Fourier-PGARCH model or Fourier-PACD model with substantially fewer parameters.

In financial contexts, sudden spikes or sharp drops in data are common, especially during events such as geopolitical tensions, regulatory shifts, or impactful market news. These abrupt changes mean that the underlying parameters often do not vary smoothly over time. As a result, a greater number of Fourier coefficients is required to represent the parameters sufficiently. This is because, even though the changes are local, all Fourier components are affected to some extent.

\section{Wavelet Analysis of Parameter Vectors}
Wavelet analysis has emerged as a powerful alternative to Fourier analysis, particularly for applications where time localization is essential. It has been widely used in areas such as statistical estimation (see \citeauthor{ ghanbari2023wavelet}, \citeyear{ ghanbari2023wavelet};  \citeauthor{karamikabir2024location}, \citeyear{karamikabir2024location}), signal denoising (\citeauthor{sahoo2024optimal}, \citeyear{sahoo2024optimal}), data compression (\citeauthor{thomas2023novel}, \citeyear{thomas2023novel}), and modeling of long memory processes (\citeauthor{de2024maximum}, \citeyear{de2024maximum}). This implies that the widely used Fourier analysis is not without certain limitations. One key limitation of Fourier analysis is its lack of time resolution. It provides information about the frequency content of a signal, but not when those frequencies occur. This makes it unsuitable for analyzing non-stationary signals, where frequency characteristics may vary over time. Wavelet analysis overcomes this issue by offering both time and frequency resolution. Unlike the sine and cosine functions used in Fourier analysis, which extend infinitely along the time axis, wavelets are localized functions—they usually exist only within a finite interval. This localization means that when a signal undergoes a local change, only a limited number of wavelet coefficients are affected, in contrast to the global effect seen in Fourier coefficients (see \citeauthor{ pewav2000}, \citeyear{ pewav2000}). As a result, wavelet methods can represent signals with local variations more efficiently, often requiring fewer coefficients than Fourier-based methods. Among the various wavelet transform techniques, we employ the orthogonal discrete wavelet transform (DWT). This choice is motivated by the projection theorem (\citeauthor{ brockwell1991time}, \citeyear{ brockwell1991time}), from which we infer that orthogonal approximations minimize mean square error. Another advantage of the DWT, particularly in relation to the objective of developing parsimonious models, is its ability to redistribute the energy or variability present in a sequence (\citeauthor{vidakovic1999statistical}, \citeyear{vidakovic1999statistical}). As a result, only a small number of wavelet coefficients are often sufficient to capture the essential structure of the sequence. For a more comprehensive discussion on the use of wavelets in statistical analysis, one can refer to \citeauthor{vidakovic1999statistical} (\citeyear{vidakovic1999statistical}), \citeauthor{pewav2000} (\citeyear{ pewav2000}), and \citeauthor{nason2008wavelet} (\citeyear{nason2008wavelet}).

If the size of $\textbf{X}$ is $M$ satisfying $M = 2^J$ for some positive integer $J$, then its DWT is obtained by (\citeauthor{pewav2000}, \citeyear{pewav2000}) 
\begin{equation}
    \mathbf{W}_{\textbf{X}} = \mathcal{W}\mathbf{X} = \begin{bmatrix}
    {V}_{\textbf{X},J}\\ 
    \mathbf{W}_{\textbf{X},1}\\
    \vdots\\
    \mathbf{W}_{\textbf{X},J}
    \end{bmatrix} = 
    \begin{bmatrix}
        W_{\textbf{X},0}\\
        W_{\textbf{X},1}\\
        \vdots\\
        W_{\textbf{X},M-1}
    \end{bmatrix},
\end{equation}
where $\mathcal{W} \in \mathbb{R}^{M\times M}$ is an orthogonal matrix associated with the DWT, $V_{\mathbf{X},J} = W_{\mathbf{X},0} = \sqrt{M}\left(\frac{1}{M}\sum\limits_{i = 1}^MX_i\right)$, and for $1 \leq j \leq J$, the set $\mathbf{W}_{\mathbf{X},j}$ consists of $2^{j-1}$ wavelet coefficients that capture changes at the resolution level $M/2^{j-1}$. Of the many orthogonal discrete wavelet transforms available in the literature, the simplest one is the Daubechies extremal phase wavelet with one vanishing moment, viz., Haar. For instance, if $\mathbf{X} = (X_0,\dotsc,X_7)^T$, its Haar DWT is
\begin{equation}\label{DWTeg}
    \begin{bmatrix}
        \frac{1}{2\sqrt{2}}(X_0+\dotsc+X_7)\\
        \frac{1}{2\sqrt{2}}(X_0+\dotsc+X_3-X_4-\dotsc-X_7)\\
        \frac{1}{2}(X_0+X_1-X_2-X_3)\\
        \frac{1}{2}(X_4+X_5-X_6-X_7)\\
        \frac{1}{\sqrt{2}}(X_0-X_1)\\
        \frac{1}{\sqrt{2}}(X_2-X_3)\\
        \frac{1}{\sqrt{2}}(X_4-X_5)\\
        \frac{1}{\sqrt{2}}(X_6-X_7)\\
    \end{bmatrix}.
\end{equation}
In this context, $\mathcal{W}$ is described as defined in \eqref{Haarmat}. The discrete wavelet transform's ability to capture local features is apparent from \eqref{DWTeg}. Given $\mathbf{W}_{\mathbf{X}}$, the original vector $\mathbf{X}$ can be obtained by applying the inverse discrete wavelet transform
\begin{equation*}
    \mathbf{X} = \mathcal{W}^T\mathbf{W}_{\mathbf{X}}.
\end{equation*}
Therefore, like the Fourier transform, the DWT of $\mathbf{X}$ offers an alternative perspective on the same vector.

Wavelet techniques have been successfully employed to reduce the number of parameters in the PARMA model, resulting in improved forecasting accuracy (\citeauthor{davis2024wavelet}, \citeyear{davis2024wavelet}). Motivated by this success, we propose the use of wavelet-based methods to represent the parameters of PGARCH and PACD models with a reduced set of statistically relevant wavelet coefficients. Since financial time series often exhibit abrupt and localized changes, the wavelet-based method is expected to achieve greater parsimony and forecasting efficiency than the Fourier-based technique in the context of PGARCH and PACD models.

If \eqref{Xasynorm} is satisfied, Theorem \ref{A.broc.asymnormaltheorem} implies that the discrete wavelet transform of the parameter vector has the property 
\begin{equation}
    N^{1/2}(\hat{\mathbf{W}}_{\mathbf{X}}-\mathbf{W}_{\mathbf{X}}) \xrightarrow{d} \mathcal{N}(\boldsymbol{0},R_{\mathbf{W}_{\mathbf{X}}}),
\end{equation} 
with $R_{\mathbf{W}_{\mathbf{X}}} = \mathcal{W}\Gamma_{\mathbf{X}}\mathcal{W}^T$. To determine which wavelet coefficients are statistically significant, we consider a hypothesis testing approach analogous to the one formulated for identifying significant Fourier coefficients. Here, the null hypothesis, denoted as $H_0: \mathbf{W}_{\mathbf{X}} = (W_{\mathbf{X}, 0},0\dotsc,0)^T$, asserts that $X_i = \sqrt{M}\left(\frac{1}{M}\sum\limits_{\ell=1}^MX_{\ell}\right), 0 \leq i \leq \nu-1$. When all parameter vectors are reduced to constant vectors, the model is simplified to the traditional GARCH or ACD model. According to 
\begin{equation}
    N^{1/2}(\hat{W}_{\mathbf{X},i}-W_{\mathbf{X},i}) \xrightarrow{d} \mathcal{N}(0,[R_{\mathbf{W}_{\mathbf{X}}}]_{ii}),
\end{equation} 
the wavelet coefficients are asymptotically normal. In this context, the $i^{th}$ entry of vector $\mathbf{W}_{\mathbf{X}}$ is labeled as $W_{\mathbf{X},i}$, while $[R_{\mathbf{W}_{\mathbf{X}}}]_{ii}$ corresponds to the diagonal elements of matrix $R_{\mathbf{W}_{\mathbf{X}}}$. The associated hypothesis test involves the null and alternative hypotheses represented by $H_0: |W_{\mathbf{X},i}|=0$ and $H_1: |W_{\mathbf{X},i}|\neq 0,$ for $1 \leq i \leq \nu-1$, respectively. The corresponding test statistic is presented by
\begin{equation}
    Z_{\mathbf{W},i} = \frac{\hat{W}_{\mathbf{X},i}}{\sqrt{[R_{\mathbf{W}_{\mathbf{X}}}]_{ii}/N}}, \quad 1 \leq i \leq \nu-1.
\end{equation}
Since Bonferroni correction is used, when $|Z_{\mathbf{W},i}|>\Phi^{-1}(\alpha/(2/(\nu-1)))$, the null hypothesis is rejected, indicating that the concerned wavelet coefficient is statistically significant and should be retained. Conversely, non-significant coefficients are set to zero. Since financial time series often contain sudden shifts, we expect that only a limited subset of the wavelet coefficients will be necessary, yielding a parsimonious wavelet-PGARCH model or wavelet-PACD model.

\begin{remark}\label{remark.per.ext}
Since the discrete wavelet transform can be applied only to vectors of size $2^J$, the vectors are periodically extended to the nearest integer of the form $2^J$ along with their corresponding asymptotic variance-covariance matrices.
\end{remark}

\section{Simulation Studies}
The ideas proposed in the previous sections are demonstrated through simulation studies in the following subsections. 
\subsection{Simulation of Fourier-PGARCH model}
Consider the $\text{Fourier-PGARCH}_{7}(1,1)$ model,
\begin{equation}
    \begin{split}
        y_t &= \epsilon_t \sqrt{h_t}\\
        \text{and} \quad h_t &= \omega_t+\alpha_ty_{t-1}^2+\beta_th_{t-1}, \, t=0,1,\dotsc,
    \end{split}
\end{equation}
satisfying (i) $\omega_t =\omega_{t+7k},k \in \mathbb{Z}$, (ii) $\alpha_t =\alpha_{t+7k},k \in \mathbb{Z}$, and (iii) $\beta_t =\beta_{t+7k},k \in \mathbb{Z}$ with Fourier representation of $\boldsymbol{\omega}$, $\boldsymbol{\alpha}$, and $\boldsymbol{\beta}$ described by
\begin{equation}
    \begin{split}
        \omega_t &= 0.7+0.45\sin \left(\frac{2\pi t}{7}\right),\\
\alpha_t &= 0.6+0.15\sin \left(\frac{2\pi t}{7} \right),\\
\text{and} \quad \beta_t &= 0.35+0.2\sin\left(\frac{2\pi t}{7} \right), \, t=0,1,\dotsc,6,
    \end{split}
\end{equation}
respectively. Thus, the model consists of only 6 parameters. Here, $\{\epsilon_t\}$ follows generalized error distribution with mean 0 and unit variance, whose probability density function is given by (\citeauthor{wurtz2006parameter}, \citeyear{wurtz2006parameter})
\begin{equation}\label{ged}
    f(\epsilon_t) = \frac{\upsilon}{\kappa_{\upsilon}2^{1+\frac{1}{\upsilon}}\Gamma(\frac{1}{\upsilon})}e^{-\frac{1}{2}|\frac{\epsilon_t}{\kappa_{\upsilon}}|^{\upsilon}}, \quad -\infty < \epsilon_t < \infty,
\end{equation}
where $\kappa_{\upsilon} = (2^{-2/\upsilon}\Gamma(1/\upsilon)/\Gamma(3/\upsilon))^{1/2}$ and $\upsilon>0$. Here, the value of $\upsilon$ was taken to be 1.8.
To study the performance of the model, 4396 observations were generated, viz. $y_0, \dotsc, y_{4395}$. The initial values $y_{-1}$ and $h_{-1}$ were assumed to be $y_{398}$ and $h_{398}$, respectively, and the first 399 observations were removed. For convenience, the remaining 3997 observations $y_{399} \dotsc y_{4395}$ were reindexed as $y_0,\dotsc,y_{3996}$. Subsequently, to estimate the model, the initial $570 \times 7 = 3990$ observations were considered, whereas the last 7 values were allocated for forecast analysis. For performing the nonlinear optimization algorithm to obtain the QMLE described in \eqref{PGARCHQMLE}, the starting values of the parameters were set to their true values. The Fourier estimates were then derived as described in Section 2. The entire process was implemented over 100 replications. $\text{Fourier-PGARCH}_{7}(1,1)$ estimates and the corresponding root mean square errors (RMSEs) are reported in Table \ref{FourierPGARCHsimres}.
\begin{table}
\caption{Estimation Results for $\text{Fourier-PGARCH}_{7}(1,1)$.}
\begin{tabular}{cccccccccccc}
\toprule
  & \multicolumn{3}{c}{$\mathbf{f}_{\boldsymbol{\omega}}$} &  & \multicolumn{3}{c}{$\mathbf{f}_{\boldsymbol{\alpha}}$} &  & \multicolumn{3}{c}{$\mathbf{f}_{\boldsymbol{\beta}}$} \\ \cmidrule{2-4} \cmidrule{6-8} \cmidrule{10-12} 
  & True           & Estimate           & RMSE             &  & True           & Estimate           & RMSE             &  & True           & Estimate           & RMSE            \\
0 & 0.70           & 0.6946             & 0.0618           &  & 0.60           & 0.5998             & 0.0348           &  & 0.35           & 0.3553             & 0.0263          \\
1 & 0.00           & -0.0057            & 0.0404           &  & 0.00           & 0.0000             & 0.0000           &  & 0.00           & 0.0000             & 0.0000          \\
2 & 0.45           & 0.4011             & 0.2146           &  & 0.15           & 0.1159             & 0.1112           &  & 0.20           & 0.2013             & 0.0503          \\
3 & 0.00           & 0.0082             & 0.0581           &  & 0.00           & -0.0040            & 0.0285           &  & 0.00           & 0.0000             & 0.0000          \\
4 & 0.00           & 0.0000             & 0.0000           &  & 0.00           & 0.0000             & 0.0000           &  & 0.00           & 0.0000             & 0.0000          \\
5 & 0.00           & 0.0000             & 0.0000           &  & 0.00           & 0.0000             & 0.0000           &  & 0.00           & 0.0000             & 0.0000          \\
6 & 0.00           & 0.0000             & 0.0000           &  & 0.00           & 0.0000             & 0.0000           &  & 0.00           & 0.0000             & 0.0000 \\
\bottomrule
\end{tabular}
\label{FourierPGARCHsimres}
\end{table}
Let $y_{t,p}$ denote the forecast value of $y_t$. The forecast performance of the models was compared using the root mean square forecast error, $\text{RMSFE} = \sqrt{\frac{1}{T}\sum\limits_{t=1}^T (y_t-y_{t,p})^2}$ and the mean absolute forecast error, $\text{MAFE} = \frac{1}{T}\sum\limits_{t=1}^T |y_t-y_{t,p}|$, where $T$ denotes the number of elements in the forecast set. For out-sample observations, the $\text{Fourier-PGARCH}_7(1,1)$ model showed a forecast accuracy \textbf{gain} of 1.16\% and 0.29\% (average RMSFE and MAFE) over the $\text{PGARCH}_7(1,1)$ model. These findings suggest that the $\text{Fourier-PGARCH}_7(1,1)$ model has a forecast efficiency superior to the $\text{PGARCH}_{7}(1,1)$ model.

\subsection{Simulation of Fourier-PACD model}
The $\text{Fourier-PACD}_7(1,1)$ model is described by
\begin{equation}
          \begin{split}
              u_t &= \psi_t \xi_t\\
          \text{and} \quad  
              \psi_t &= \lambda_t + \gamma_t u_{t-1} + \delta_t \psi_{t-1}, \, t=0,1,\dotsc,
              \end{split}
          \end{equation}
          with (i) $\lambda_t = \lambda_{t+7k}, k \in \mathbb{Z} $, (ii) $\gamma_t = \gamma_{t+7k}, k \in \mathbb{Z}$, (iii) $\delta_t = \delta_{t+7k}, k \in \mathbb{Z} $, and $\sigma_t^2 = \sigma_{t+7k}^2, k \in \mathbb{Z}$, where the Fourier representation of $\boldsymbol{\lambda}, \boldsymbol{\gamma}, \boldsymbol{\delta}$ and $\boldsymbol{\sigma^2}$ is specified as 
\begin{equation}
    \begin{split}
 \lambda_t &= 0.55+0.45\cos \left(\frac{2\pi t}{7}\right),\\
  \gamma_t &= 0.65+0.14\cos \left(\frac{2\pi t}{7}\right),\\
    \delta_t &= 0.32+0.18\cos \left(\frac{2\pi t}{7}\right),\\
  \text{and} \quad  \sigma_t^2 &= 0.4+0.3\cos \left(\frac{2\pi t}{7}\right), \, t=0,1,\dotsc,6,
    \end{split}
\end{equation} 
respectively. Hence, the model includes only eight parameters. Here, $\{\xi_t\}$ follows the gamma distribution with unit mean and variance $\sigma_t^2$, whose probability density function is 
\begin{equation}\label{gammadist}
f(\xi_t) = \frac{(\sigma_t^{-2})^{\sigma_t^{-2}}}{\Gamma(\sigma_t^{-2})} e^{\sigma_t^{-2}\xi_t}\xi_t^{\sigma_t^{-2}-1}, \quad \xi_t > 0, \sigma_t>0.
\end{equation}
A total of 2191 data points were simulated from the model, viz. $u_0,\dotsc,u_{2190}$. The initial values $u_{-1}$ and $\psi_{-1}$ were set as $u_{195}$ and $h_{195}$, respectively, and the first 196 observations were discarded. For simplicity, the remaining 1995 observations $u_{196},\dotsc,u_{2190}$ were relabeled as $u_0,\dotsc,u_{1994}$. The first $284 \times 7= 1988$ values were utilized for estimation, while the remaining 7 observations were reserved to assess forecast performance. To carry out the nonlinear optimization algorithm for estimating the QMLE in \eqref{PACDQMLE}, the true parameter values were used as starting points. The Fourier estimates were obtained following the procedure detailed in Section 2. The entire process was implemented across 100 replications. The estimates of the $\text{Fourier-PACD}_7(1,1)$ model along with their associated root mean square errors (RMSEs) are presented in Table \ref{Fourierpacdsimres}. 
\begin{table}
\caption{Estimation Results for $\text{Fourier-PACD}_7(1,1)$.}
\begin{tabular}{cccccccc}
\toprule
  & \multicolumn{3}{c}{$\mathbf{f}_{\boldsymbol{\lambda}}$}             &  & \multicolumn{3}{c}{$\mathbf{f}_{\boldsymbol{\gamma}}$}                \\
 \cmidrule{2-4} \cmidrule{6-8}
  & True           & Estimate           & RMSE             &  & True            & Estimate            & RMSE             \\
0 & 0.55           & 0.5415             & 0.0842           &  & 0.65            & 0.6489              & 0.0306           \\
1 & 0.45           & 0.3557             & 0.2448           &  & 0.14            & 0.1084              & 0.0894           \\
2 & 0.00           & 0.0036             & 0.0360           &  & 0.00            & -0.0012             & 0.0123           \\
3 & 0.00           & 0.0000             & 0.0000           &  & 0.00            & 0.0026              & 0.0184           \\
4 & 0.00           & 0.0000             & 0.0000           &  & 0.00            & 0.0013              & 0.0126           \\
5 & 0.00           & 0.0000             & 0.0000           &  & 0.00            & 0.0016              & 0.0163           \\
6 & 0.00           & 0.0000             & 0.0000           &  & 0.00            & -0.0010             & 0.0096           \\
  & \multicolumn{3}{c}{$\mathbf{f}_{\boldsymbol{\delta}}$} &  & \multicolumn{3}{c}{$\mathbf{f}_{\boldsymbol{\sigma}^2}$} \\
   \cmidrule{2-4} \cmidrule{6-8}
0 & 0.32           & 0.3236             & 0.0317           &  & 0.40             & 0.3939              & 0.0184           \\
1 & 0.18           & 0.1728             & 0.0684           &  & 0.30             & 0.2966              & 0.0259           \\
2 & 0.00           & -0.0001            & 0.0190           &  & 0.00            & 0.0000              & 0.0000           \\
3 & 0.00           & 0.0000             & 0.0000           &  & 0.00            & 0.0000              & 0.0000           \\
4 & 0.00           & 0.0013             & 0.0129           &  & 0.00            & 0.0000              & 0.0000           \\
5 & 0.00           & -0.0018            & 0.0179           &  & 0.00            & 0.0000              & 0.0000           \\
6 & 0.00           & 0.0001             & 0.0148           &  & 0.00            & 0.0000              & 0.0000 \\
\bottomrule
\end{tabular}
\label{Fourierpacdsimres}
\end{table}

When compared to the $\text{PGARCH}_7(1,1)$ model, the Fourier-based model demonstrated an increase in forecast efficiency of 2.49\% in RMSFE and 0.26\% in MAFE on the out-of-sample observations. Thus, the $\text{Fourier-PACD}_7(1,1)$ model outperforms the $\text{PACD}_7(1,1)$ model in terms of forecasting efficiency.
\subsection{Simulation of wavelet-PGARCH model}
The $\text{wavelet-PGARCH}_8(1,1)$ model is specified as
\begin{equation}
    \begin{split}
        y_t &= \epsilon_t\sqrt{h_t}\\
        \text{and} \quad h_t &=\omega_t+\alpha_ty_{t-1}^2+\beta_th_{t-1}, \, t=0,1,\dotsc,
    \end{split}
\end{equation}
satisfying the conditions (i) $\omega_t =\omega_{t+8k},k \in \mathbb{Z}$, (ii) $\alpha_t =\alpha_{t+8k},k \in \mathbb{Z}$, and (iii) $\beta_t =\beta_{t+8k},k \in \mathbb{Z}$ where DWT of $\boldsymbol{\omega}, \boldsymbol{\alpha}$, and $\boldsymbol{\beta}$ are $\mathbf{W}_{\boldsymbol{\omega}} = (2,1,0\dotsc,0)^T$, $\mathbf{W}_{\boldsymbol{\alpha}} = (1.9,0.5,0\dotsc,0)^T$, and $\mathbf{W}_{\boldsymbol{\beta}} = (0.85,0.35,0\dotsc,0)^T$, respectively. Here, $\{\epsilon_t\}$ follows generalized error distribution as given in \eqref{ged}. The model, in this case, comprises of only 6 parameters. To evaluate the model's capability, a total of 4400 observations were simulated, viz. $y_0,\dotsc,y_{4399}$. The intial values $y_{-1}$ and $h_{-1}$ were assumed to be $y_{399}$ and $h_{399}$, respectively, and the first 400 observations were removed. For convenience, the remaining 4000 observations $y_{400},\dotsc,y_{4399}$ were reindexed as $y_0,\dotsc,y_{3999}$. Of these, $499 \times 8 = 3992$ were employed for parameter estimation, while the final 8 were reserved for assessing forecast performance. For performing the nonlinear optimization algorithm to obtain the QMLE described in \eqref{PGARCHQMLE}, the starting values of the parameters were set to their true values. The wavelet estimates were then derived as described in Section 3. The wavelet used here was the Daubechies extremal phase wavelet having eight vanishing moments, referred to as D(8), where \eqref{D8mat} is its transform matrix. The discrete wavelet transform was performed using the \texttt{wavethresh} package in R, as described in \citeauthor{wavethresh} (\citeyear{wavethresh}). The entire process was implemented across 100 replications. Table \ref{waveletPGARCHsimres} presents the results of the D(8) discrete wavelet transform analysis. 
\begin{table}
\caption{Estimation Results for $\text{wavelet-PGARCH}_8(1,1)$.}
\begin{tabular}{cccccccccccc}
\toprule
  & \multicolumn{3}{c}{$\mathbf{W}_{\boldsymbol{\omega}}$} &  & \multicolumn{3}{c}{$\mathbf{W}_{\boldsymbol{\alpha}}$} &  & \multicolumn{3}{c}{$\mathbf{W}_{\boldsymbol{\beta}}$} \\ \cmidrule{2-4} \cmidrule{6-8} \cmidrule{10-12} 
  & True           & Estimate           & RMSE             &  & True           & Estimate           & RMSE             &  & True           & Estimate           & RMSE            \\
0 & 2.00           & 1.9577             & 0.1920           &  & 1.90           & 1.9163             & 0.1272           &  & 0.85           & 0.8608             & 0.0908          \\
1 & 1.00           & 0.9241             & 0.3429           &  & 0.50           & 0.5072             & 0.1275           &  & 0.35           & 0.3195             & 0.1568          \\
2 & 0.00           & -0.0085            & 0.0851           &  & 0.00           & 0.0032             & 0.0656           &  & 0.00           & -0.0029            & 0.0502          \\
3 & 0.00           & 0.0000             & 0.0000           &  & 0.00           & -0.0038            & 0.0656           &  & 0.00           & 0.0022             & 0.0224          \\
4 & 0.00           & 0.0000             & 0.0000           &  & 0.00           & -0.0089            & 0.0633           &  & 0.00           & 0.0000             & 0.0000          \\
5 & 0.00           & 0.0000             & 0.0000           &  & 0.00           & -0.0030            & 0.0299           &  & 0.00           & -0.0031            & 0.0305          \\
6 & 0.00           & 0.0000             & 0.0000           &  & 0.00           & 0.0000             & 0.0000           &  & 0.00           & 0.0000             & 0.0000          \\
7 & 0.00           & 0.0000             & 0.0000           &  & 0.00           & -0.0002            & 0.0433           &  & 0.00           & 0.0020             & 0.0197 \\
\bottomrule
\end{tabular}
\label{waveletPGARCHsimres}
\end{table}
Compared to the $\text{PGARCH}_8(1,1)$ model, the $\text{wavelet-PGARCH}_8(1,1)$ model achieved a forecast efficiency \textbf{gain} of 2.11\% and 1.10\% based on the average RMSFE and MAFE. These results highlight the superior forecasting performance of the proposed $\text{wavelet-PGARCH}_8(1,1)$ model.

\subsection{Simulation of wavelet-PACD model}
The $\text{wavelet-PACD}_8(1,1)$ model is formulated as  
\begin{equation}
          \begin{split}
              u_t &= \psi_t \xi_t\\
          \text{and} \quad  
              \psi_t &= \lambda_t + \gamma_t u_{t-1} + \delta_t \psi_{t-1}, \, t=0,1,\dotsc,
              \end{split}
          \end{equation}
         subject to conditions (i) $\lambda_t = \lambda_{t+8k}, k \in \mathbb{Z} $, (ii) $\gamma_t = \gamma_{t+8k}, k \in \mathbb{Z}$, (iii) $\delta_t = \delta_{t+8k}, k \in \mathbb{Z} $, and $\sigma_t^2 = \sigma_{t+8k}^2, k \in \mathbb{Z}$, where the discrete wavelet transforms (DWT) of the vectors $\boldsymbol{\lambda}, \boldsymbol{\gamma}, \boldsymbol{\delta}$, and $\boldsymbol{\sigma^2}$ are denoted by $\mathbf{W}_{\boldsymbol{\lambda}} = (1.9,1.2,0,\dotsc,0)^T, \mathbf{W}_{\boldsymbol{\gamma}} = (2.0, 0.3,0,\dotsc,0)^T, \mathbf{W}_{\boldsymbol{\delta}} = (0.84,0.40,0\dotsc,0)^T$, and $\mathbf{W}_{\boldsymbol{\sigma}^2} = (1.20,0.70,0,\dotsc,0)^T$, respectively. Hence, the model consists of only eight parameters. Here, $\{\xi_t\}$ follows gamma distribution as given in \eqref{gammadist}. To analyze the performance of the model, 2200 data points were generated, viz. $u_0,\dotsc,u_{2199}$. The initial values $u_{-1}$ and $\psi_{-1}$ were set as $u_{199}$ and $\psi_{199}$, respectively, and the first 200 observations were discarded. For simplicity, the remaining 2000 observations $u_{200},\dotsc,u_{2199}$ were relabeled as $u_0,\dotsc,u_{1999}$.  Among these, the first $ 249 \times 8 = 1992 $ observations were used for parameter estimation, and the remaining 8 values were set aside for forecast evaluation. To carry out the nonlinear optimization algorithm for estimating the QMLE in \eqref{PACDQMLE}, the true parameter values were used as starting points. The wavelet estimates were obtained following the procedure detailed in Section 3. The wavelet applied here was the Daubechies extremal phase wavelet with five vanishing moments, denoted D(5), and its corresponding transformation matrix is \eqref{D5mat}.  The entire process was implemented across 100 replications. The D(5) DWT estimates and the corresponding RMSE are reported in Table \ref{wavpacdsimres}. 
         \begin{table}
         \caption{Estimation Results for $\text{wavelet-PACD}_7(1,1)$.}
\begin{tabular}{cccccccc}
\toprule
  & \multicolumn{3}{c}{$\mathbf{W}_{\boldsymbol{\lambda}}$} &  & \multicolumn{3}{c}{$\mathbf{W}_{\boldsymbol(\gamma)}$}   \\
   \cmidrule{2-4} \cmidrule{6-8}
  & True            & Estimate           & RMSE             &  & True            & Estimate            & RMSE             \\
0 & 1.90            & 1.9424             & 0.3048           &  & 2.00            & 2.0113              & 0.0983           \\
1 & 1.20            & 1.1061             & 0.4899           &  & 0.30            & 0.2633              & 0.1749           \\
2 & 0.00            & -0.0047            & 0.0467           &  & 0.00            & 0.0038              & 0.0267           \\
3 & 0.00            & 0.0147             & 0.1475           &  & 0.00            & 0.0000              & 0.0000           \\
4 & 0.00            & -0.0107            & 0.1067           &  & 0.00            & -0.0016             & 0.0162           \\
5 & 0.00            & 0.0000             & 0.0000           &  & 0.00            & 0.0000              & 0.0000           \\
6 & 0.00            & 0.0000             & 0.0000           &  & 0.00            & 0.0035              & 0.0354           \\
7 & 0.00            & 0.0000             & 0.0000           &  & 0.00            & -0.0034             & 0.0339           \\
  & \multicolumn{3}{c}{$\mathbf{W}_{\boldsymbol{\delta}}$}  &  & \multicolumn{3}{c}{$\mathbf{W}_{\boldsymbol{\sigma^2}}$} \\
   \cmidrule{2-4} \cmidrule{6-8}
0 & 0.84            & 0.8312             & 0.0895           &  & 1.20            & 1.1782              & 0.0538           \\
1 & 0.40            & 0.3945             & 0.0903           &  & 0.70            & 0.6852              & 0.0589           \\
2 & 0.00            & 0.0000             & 0.0000           &  & 0.00            & -0.0014             & 0.0100           \\
3 & 0.00            & 0.0000             & 0.0000           &  & 0.00            & -0.0020             & 0.0201           \\
4 & 0.00            & 0.0015             & 0.0147           &  & 0.00            & 0.0000              & 0.0000           \\
5 & 0.00            & -0.0016            & 0.0163           &  & 0.00            & 0.0000              & 0.0000           \\
6 & 0.00            & -0.0042            & 0.0421           &  & 0.00            & 0.0017              & 0.0171           \\
7 & 0.00            & 0.0028             & 0.0284           &  & 0.00            & -0.0060             & 0.0597     \\
\bottomrule
\end{tabular}
\label{wavpacdsimres}
\end{table}

For out-of-sample observations, the $\text{wavelet-PACD}_7(1,1)$ model showed a forecast accuracy \textbf{gain} of 1.05\% and 1.35\% (average RMSFE and MAFE) over the $\text{PACD}_7(1,1)$ model. Thus, the $\text{wavelet-PACD}_7(1,1)$ model outperforms the $\text{PACD}_7(1,1)$ model in terms of forecast precision.

\section{Data Analysis}
The practical applicability of the proposed ideas is demonstrated through two data analysis studies.
\subsection{Data Analysis - 1}
\begin{figure}
    \centering
    \begin{subfigure}[b]{\textwidth}
        \centering
\includegraphics[width=\textwidth, height=3.2cm]{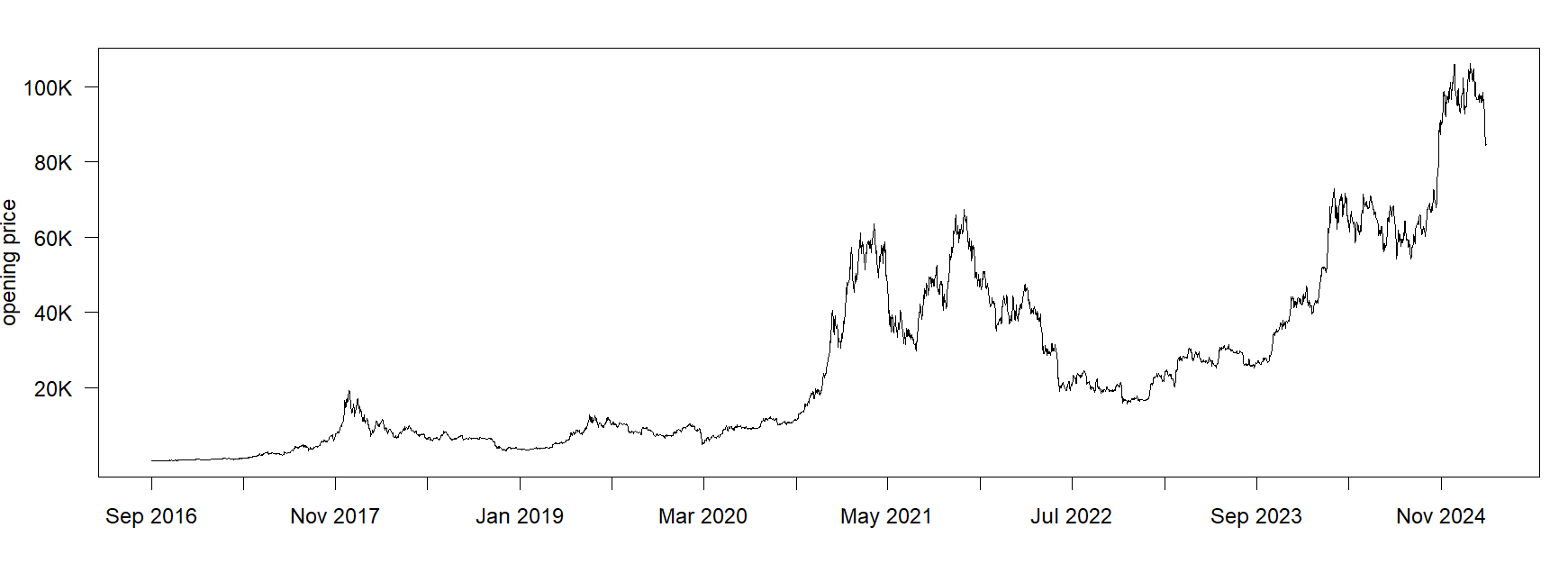}
        \caption{}
        \label{pltopenprice}
    \end{subfigure}
    \hfill
    \begin{subfigure}[b]{\textwidth}
        \centering
    \includegraphics[width=\textwidth]{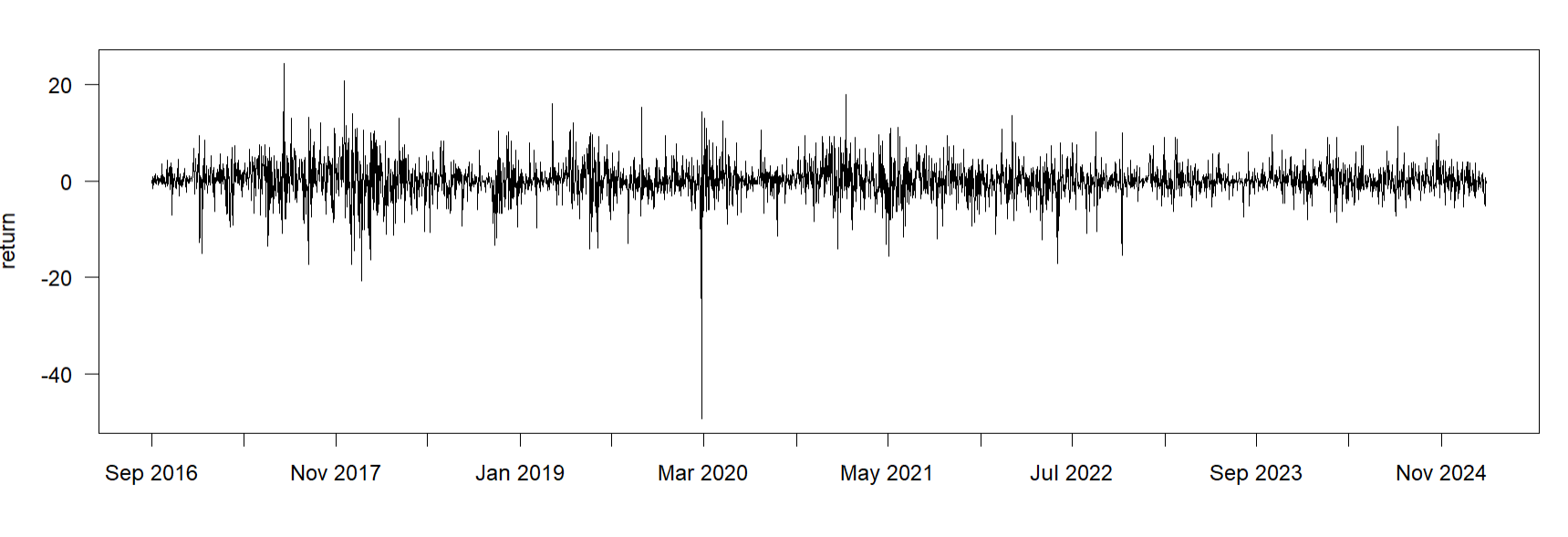}
        \caption{}
        \label{pltsqret}
    \end{subfigure}
    \caption{(a) Plot of the $p_t$ series.  (b) Plot of the $y_t$ series.}
\end{figure}

The dataset consists of the daily opening prices of Bitcoin (in USD) from September 17, 2016, to March 1, 2025, obtained from \textit{https://in.investing.com/crypto/bitcoin/btc-usd-historical-data}. Let $p_t, t = 0, \dotsc,3087$, denote the daily opening price. From this series, the nominal returns $y_t$ (in percentage) were computed using (\citeauthor{bollerslev1996periodic}, \citeyear{bollerslev1996periodic})
\begin{equation}
y_t = 100[\log(p_t)-\log(p_{t-1})], \, t = 1, \dotsc, 3087.
\end{equation}
For convenience, the series $\{y_t\}$ was reindexed with $t$ ranging from 0 to 3086. Previous studies, including \citeauthor{dorfleitner2018cryptocurrencies} (\citeyear{dorfleitner2018cryptocurrencies}) and \citeauthor{aharon2019bitcoin} (\citeyear{aharon2019bitcoin}), have reported a day-of-the-week effect in Bitcoin returns. Time series plots of $p_t$ and $y_t$ are presented in Figures \ref{pltopenprice} and \ref{pltsqret}, respectively. 
The descriptive statistics for both the full return series and the day-wise series are reported in Table \ref{descriptivestatisticsPGARCH}, indicating the presence of day-of-the-week patterns in returns, thereby supporting the assumption $\nu = 7$. 
\begin{table}
\centering
\caption{Descriptive Statistics.}
\begin{tabular}{ccccccc}
\toprule
Day       & Minimum   & Maximum   & Mean     & Std. Dev. & Kurtosis & Skewness \\
\midrule
Full Series & -49.2619  & 24.3483  & 0.1598   & 3.7588   & 15.8778  & -0.7739  \\
Monday      & -11.9600  & 12.0313  & 0.0428   & 2.8821   & 6.4799   & -0.3923  \\
Tuesday     & -17.0024  & 17.8685  & 0.4331   & 4.1998   & 5.3123   & 0.0542   \\
Wednesday   & -20.6384  & 16.0851  & 0.1083   & 3.6210   & 7.1903   & -0.5356  \\
Thursday    & -15.5413  & 15.2547  & 0.3024   & 3.9035   & 5.4617   & -0.2179  \\
Friday      & -49.2619  & 24.3483  & -0.1549  & 4.7445   & 30.7529  & -2.1641  \\
Saturday    & -17.2302  & 15.2324  & 0.1706   & 3.7399   & 5.6449   & -0.0324  \\
Sunday      & -14.4333  & 12.9371  & 0.2165   & 2.7756   & 8.2482   & -0.3185  \\
\bottomrule
\end{tabular}
\label{descriptivestatisticsPGARCH}
\end{table}
\begin{figure}
    \centering
    \begin{subfigure}[b]{0.45\textwidth}
        \centering
        \includegraphics[width=\textwidth, height=3.2cm]{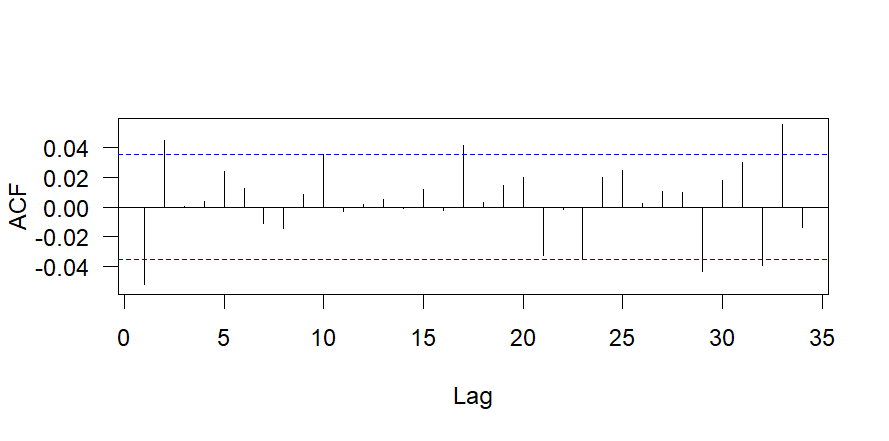}
        \caption{}
        \label{acfret}
    \end{subfigure}
    \hfill
    \begin{subfigure}[b]{0.45\textwidth}
        \centering
        \includegraphics[width=\textwidth]{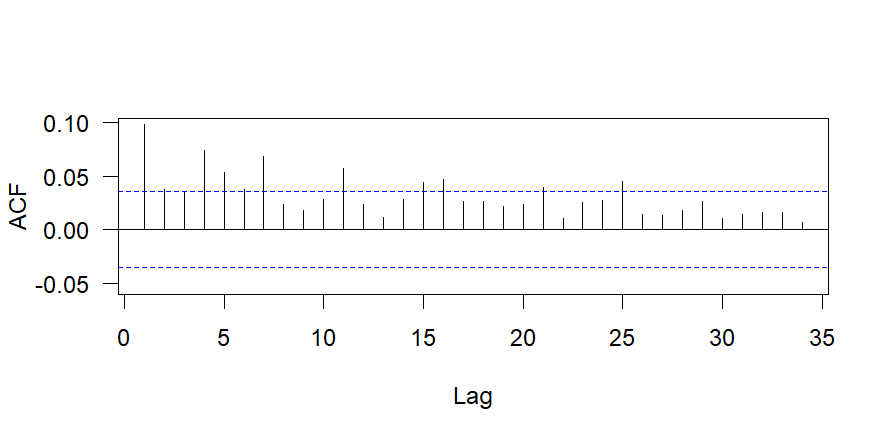}
        \caption{}
        \label{acfret2}
    \end{subfigure}
    \begin{subfigure}[b]{0.45\textwidth}
        \centering
        \includegraphics[width=\textwidth]{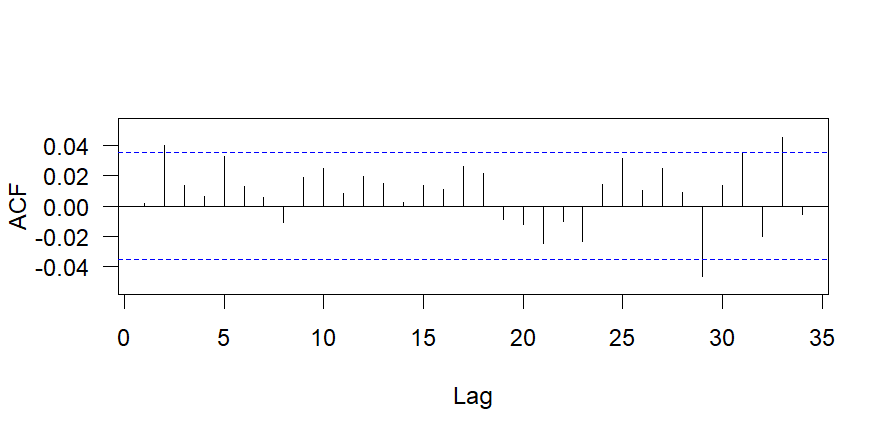}
        \caption{}
        \label{pgarchres}
    \end{subfigure}
    \hfill
    \begin{subfigure}[b]{0.45\textwidth}
        \centering
        \includegraphics[width=\textwidth]{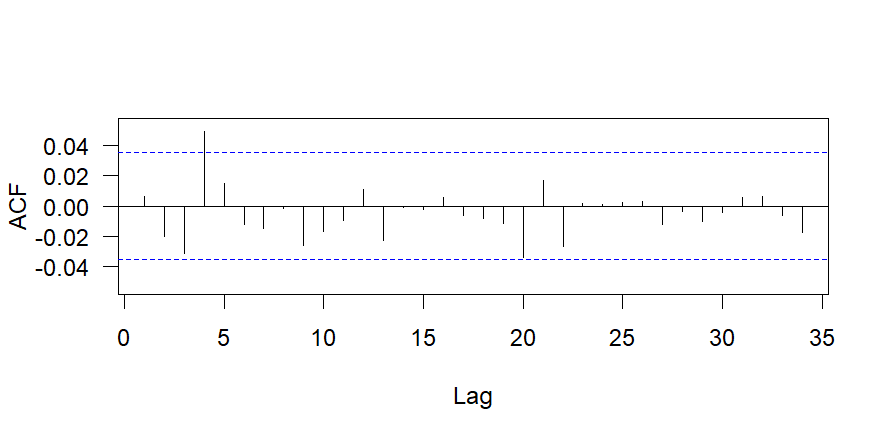}
        \caption{}
        \label{pgarchres2}
    \end{subfigure}
    \begin{subfigure}[b]{0.45\textwidth}
        \centering
        \includegraphics[width=\textwidth, height=3.2cm]{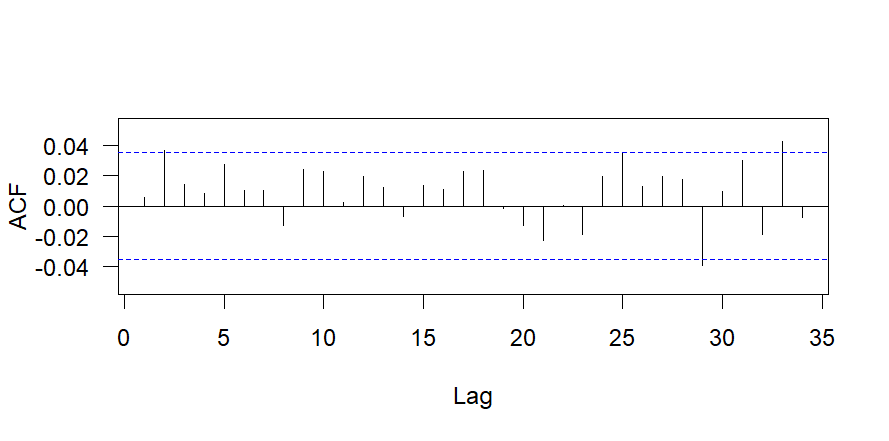}
        \caption{}
        \label{foupgarchres}
    \end{subfigure}
    \hfill
    \begin{subfigure}[b]{0.45\textwidth}
        \centering
        \includegraphics[width=\textwidth, height=3.2cm]{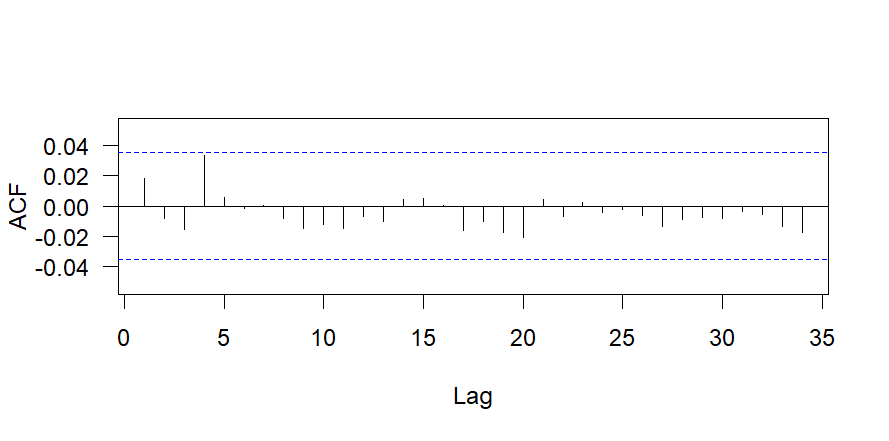}
        \caption{}
        \label{foupgarchres2}
    \end{subfigure}
    \hfill
    \begin{subfigure}[b]{0.45\textwidth}
        \centering
        \includegraphics[width=\textwidth, height=3.2cm]{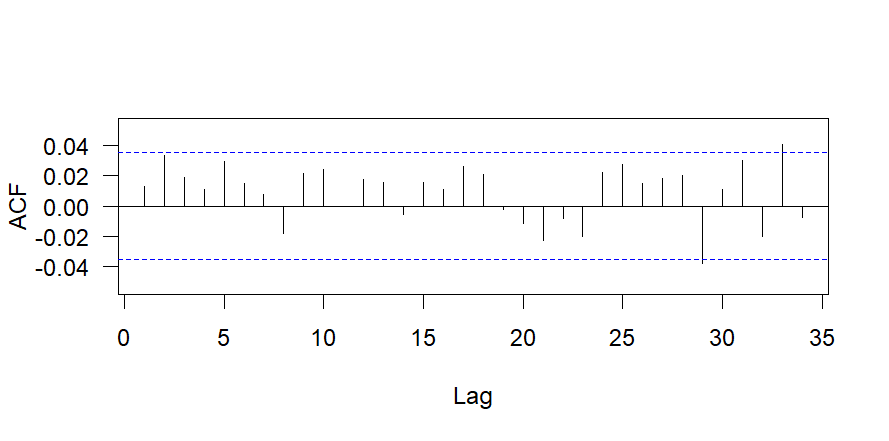}
        \caption{}
        \label{wavpgarchres}
    \end{subfigure}
    \hfill
    \begin{subfigure}[b]{0.45\textwidth}
        \centering
        \includegraphics[width=\textwidth, height=3.2cm]{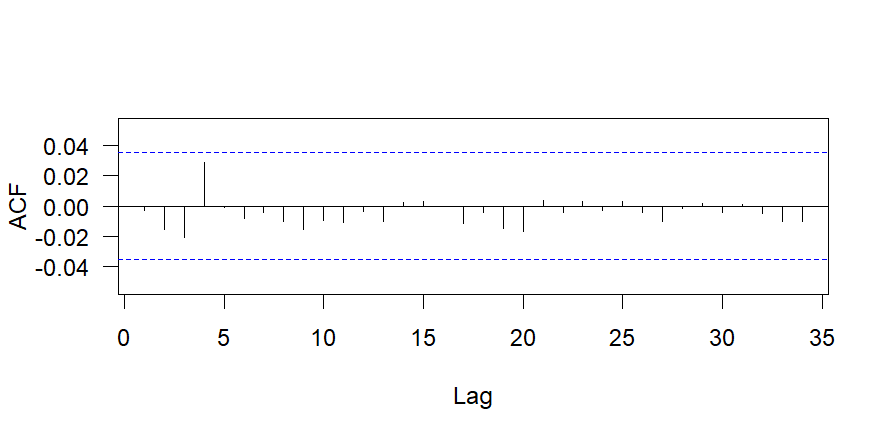}
        \caption{}
        \label{wavpgarchres2}
    \end{subfigure}
    \hfill
    \caption{(a) ACF plot of the $y_t$ series. (b) ACF plot of the $y_t^2$ series. (c) ACF plot of $\text{PGARCH}_7(1,1)$ residuals. (d) ACF plot of $\text{PGARCH}_7(1,1)$ squared residuals. (e) ACF plot of Fourier-$\text{PGARCH}_7(1,1)$ residuals. (f) ACF plot of Fourier-$\text{PGARCH}_7(1,1)$ squared residuals. (g) ACF plot of wavelet-$\text{PGARCH}_7(1,1)$ residuals. (h) ACF plot of wavelet-$\text{PGARCH}_7(1,1)$ squared residuals.}
\end{figure}
ACF plots of the returns and their squared values, displayed in Figure \ref{acfret} and Figure \ref{acfret2}, reveal mild serial dependence in returns and strong serial autocorrelation in squared returns. This observation was validated through the Ljung-Box test at various lags. For example, the p-values at lags 20 and 30 for returns are 0.0716 and 0.0229, respectively, while those for squared returns are less than $2.2 \times 10^{-16}$. Of the total of 3087 return observations, the first $440 \times 7 = 3080 $ points were used for model fitting, and the remaining 7 points were used for forecast comparison. The initial values $y_{-1}$ and $h_{-1}$ were assumed to be $y_{-1+7} = y_6$ and $y_6^2$, respectively. To compute the QMLE in \eqref{PGARCHQMLE}, starting values for the nonlinear optimization algorithm were obtained by minimizing the expression in \eqref{PGARCHQMLE} using Genetic Algorithm implemented via the \texttt{GA} package in R (see \citeauthor{scrucca2013ga}, \citeyear{scrucca2013ga}).

Figures \ref{pgarchres} and \ref{pgarchres2} display the ACF plots of the residuals and squared residuals of the $\text{PGARCH}_7(1,1)$ model. Using the Ljung-Box test, the p-values obtained at lags 20 and 30 are 0.4219 and 0.1450 for residuals, and 0.2589 and 0.5856 for squared residuals, indicating that serial autocorrelation is not statistically significant.

Table \ref{foupgarchanalysisres} displays the estimation results for $ \text{Fourier-PGARCH}_7(1,1) $. 
\begin{table}
\caption{Estimation Results for Fourier-$\text{PGARCH}_7(1,1)$.}
\begin{tabular}{cccccccccc}
\toprule
  & \multicolumn{2}{c}{$\mathbf{f}_{\boldsymbol{\omega}}$} &  & \multicolumn{2}{c}{$\mathbf{f}_{\boldsymbol{\alpha}}$} &  & \multicolumn{2}{c}{$\mathbf{f}_{\boldsymbol{\beta}}$} &  \\
   \cmidrule{2-3} \cmidrule{5-6} \cmidrule{8-9}
  & Estimate                   & Z-score                   &  & Estimate                   & Z-score                   &  & Estimate                   & Z-score                  &  \\
0 & \textbf{0.5820}                     & -                         &  & \textbf{0.1448}                     & -                         &  & \textbf{0.9160}                     & -                        &  \\
1 & -0.3755                    & -0.40                     &  & 0.0088                     & 0.34                      &  & -0.1936                    & -1.90                    &  \\
2 & 1.0189                     & 0.99                      &  & 0.1000                     & 1.95                      &  & 0.1712                     & 1.13                     &  \\
3 & -0.7590                    & -0.45                     &  & \textbf{-0.1206*}                    & -2.73                     &  & -0.3359                    & -1.60                    &  \\
4 & -0.5584                    & -0.34                     &  & 0.0479                     & 1.35                      &  & 0.0451                     & 0.24                     &  \\
5 & 0.5625                     & 0.36                      &  & 0.0195                     & 0.50                      &  & -0.0027                    & -0.01                    &  \\
6 & -0.5813                    & -0.24                     &  & -0.0256                    & -0.63                     &  & -0.2033                    & -0.75                    & \\
\bottomrule
\end{tabular}
\footnotetext{The Fourier coefficients which are statistically significant $(|Z_{\mathbf{f},i}|>2.64)$ are indicated in *. The $\text{Fourier-PGARCH}_7(1,1)$ coefficients are written in bold.}
\label{foupgarchanalysisres}
\end{table}
The estimated $ \text{Fourier-PGARCH}_7(1,1) $ model is specified by 
\begin{equation}
\begin{split}
\hat{\omega}_t &= 0.5820,\\
\hat{\alpha}_t &= 0.1448-0.1206 \cos \left(\frac{4\pi t}{7}\right),\\
\text{and} \quad \hat{\beta}_t &= 0.9160, \, t=0,1,\dotsc,6.\\
\end{split}
\end{equation}
The ACF plots for residuals and squared residuals of this model are illustrated in Figures \ref{foupgarchres} and \ref{foupgarchres2}. The Ljung-Box test yielded p-values of 0.5435 and 0.2703 at lags 20 and 30, respectively, for the residuals, and 0.9270 and 0.9965 for the squared residuals at the same lags. These results confirm the absence of significant serial correlation, suggesting that the reduced Fourier-based model, with only four parameters, is statistically adequate for modeling the data. This marks a substantial reduction from the original 21 parameters in the \( \text{PGARCH}_7(1,1) \) model.

Since the vectors $\boldsymbol{\omega}, \boldsymbol{\alpha}$, and $\boldsymbol{\beta}$ have lengths of 7, which is not a power of two required for the computation of DWT, each vector was periodically extended to a length of 8, as outlined in Remark \ref{remark.per.ext}. Let D($x$) denote the Daubechies extremal phase wavelet with $x$ vanishing moments, and LA($x$) denote the Daubechies least asymmetric wavelet with $x$ vanishing moments. Among the Daubechies extremal phase and least asymmetric wavelets available in the \texttt{wavethresh} package in R, the analysis with wavelets D(1) (Haar), D(2), D(7), LA(5), LA(8), and LA(9) resulted in $\text{wavelet-PGARCH}_7(1,1)$ models with just three parameters — in other words, the $\text{PGARCH}_7(1,1) $ model simplified to the traditional GARCH model. For instance, the results of the LA(5) DWT analysis are reported in Table \ref{wavpgarchdares}. 
\begin{table}
\caption{Estimation Results for $\text{wavelet-PGARCH}_7(1,1)$.}
\begin{tabular}{cccccccccc}
\toprule
  & \multicolumn{2}{c}{$\mathbf{W}_{\boldsymbol{\omega}}$} &  & \multicolumn{2}{c}{$\mathbf{W}_{\boldsymbol{\alpha}}$} &  & \multicolumn{2}{c}{$\mathbf{W}_{\boldsymbol{\beta}}$} &  \\
     \cmidrule{2-3} \cmidrule{5-6} \cmidrule{8-9}
  & Estimate                       & Z-score               &  & Estimate                                             & Z-score                                             &  & Estimate                   & Z-score                  &  \\
0 & \textbf{1.4439}                & -                     &  & \textbf{0.3770}                                               & -                                                   &  & \textbf{2.4027}                     & -                        &  \\
1 & -1.2754                        & -0.74                 &  & -0.1566                                              & -1.91                                               &  & -0.0805                    & -0.34                    &  \\
2 & 0.0196                         & 0.01                  &  & 0.0825                                               & 1.96                                                &  & 0.4245                     & 1.69                     &  \\
3 & -2.1081                        & -1.15                 &  & -0.0671                                              & -0.82                                               &  & -0.5390                    & -1.88                    &  \\
4 & -0.6030                        & -0.13                 &  & -0.1241                                              & -1.97                                               &  & -0.1328                    & -0.28                    &  \\
5 & 0.1619                         & 0.08                  &  & 0.0135                                               & 0.47                                                &  & 0.2069                     & 0.98                     &  \\
6 & 0.6984                         & 0.47                  &  & -0.0973                                              & -1.31                                               &  & -0.2107                    & -0.68                    &  \\
7 & 1.6917                         & 0.44                  &  & 0.2106                                               & 2.17                                                &  & 0.6681                     & 1.34                     & \\
\bottomrule
\end{tabular}
\footnotetext{The $\text{wavelet-PGARCH}_7(1,1)$ coefficients are written in bold. (None of the wavelet coefficients are statistically significant, as all the corresponding Z-scores are less than 2.69.)}
\label{wavpgarchdares}
\end{table}
The estimated DWT for $\boldsymbol{\omega}, \boldsymbol{\alpha}$ and $\boldsymbol{\beta}$ are $\hat{\mathbf{W}}_{\boldsymbol{\omega}} = (1.4439, 0, \dotsc,0)^T, \hat{\mathbf{W}}_{\boldsymbol{\alpha}}  = (0.3770,0,\dotsc,0)^T$ and $\hat{\mathbf{W}}_{\boldsymbol{\beta}}  = (2.4027,0,\dotsc,0)^T$, respectively. Note that all these models were found to be adequate for the data, since the Ljung-Box test statistics at lags 20 and 30 yield p-values of 0.4368 and 0.2327 for residuals and 0.9818 and 0.9998 for squared residuals, demonstrating the absence of significant autocorrelation. Figures \ref{wavpgarchres} and \ref{wavpgarchres2} show the ACF plots for the residuals and squared residuals of the \( \text{wavelet-PGARCH}_7(1,1) \) model. A summary of the comparison across different wavelets is provided in Table \ref{PGARCHdiffwavanalysis}.

In terms of forecast performance, the $ \text{Fourier-PGARCH}_7(1,1) $ model resulted in a forecast accuracy \textit{loss} of 91.62\% and 109.52\% (based on RMSFE and MAFE) compared to the standard \( \text{PGARCH}_7(1,1) \). In contrast, the \( \text{wavelet-PGARCH}_7(1,1) \) model achieved forecast accuracy \textbf{gains} of 7.54\% and 0.66\% in RMSFE and MAFE, respectively, relative to the original model, highlighting both its parsimony and superior forecasting ability. 
Since the $\text{PGARCH}_7(1,1)$ model was reduced to GARCH(1,1), the GARCH(1,1) model was also directly fitted to the data. Although the GARCH(1,1) model could sufficiently capture the dependency structure, as confirmed by Ljung-Box tests at the 5\% significance level, the $\text{wavelet-PGARCH}_7(1,1)$ model outperformed it in terms of forecast accuracy. Specifically, the $\text{wavelet-PGARCH}_7(1,1)$ model achieved a forecast efficiency \textbf{gain} of 0.89\% and 1.09\% in terms of RMSFE and MAFE, respectively, when compared to the GARCH(1,1) model. Thus, the $\text{wavelet-PGARCH}_7(1,1)$ model can be considered the best among all the models considered in this study.

\subsection{Data Analysis - 2}

\begin{figure}
    \centering
    \includegraphics[width=\textwidth]{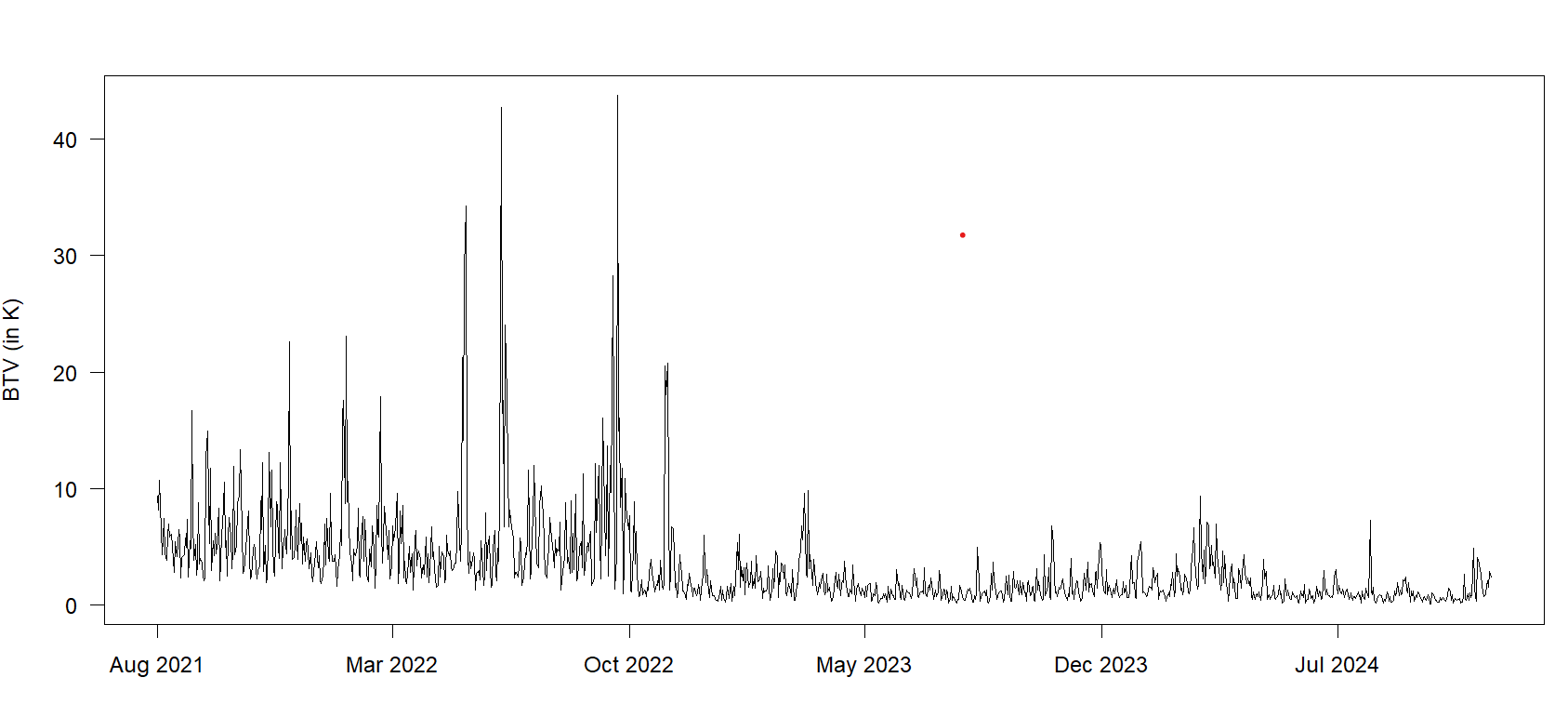}
    \caption{Plot of Bitcoin trading volume from August 7, 2021 to November 22, 2024.}
    \label{timeplotBTvolzoom}
\end{figure}

The data set consists of the daily trading volume (in 1000s) of Bitcoin against USD from August 7, 2021, to November 22, 2024, obtained from \textit{https://in.investing.com/crypto/bitcoin/btc-usd-historical-data}. The existence of a day-of-the-week effect in Bitcoin volume has been mentioned in studies like \citeauthor{wang2020time} (\citeyear{wang2020time}) and \citeauthor{aknouche2022periodic} (\citeyear{aknouche2022periodic}). The time series plot of Bitcoin trading volume is displayed in Figure \ref{timeplotBTvolzoom}. 
The descriptive statistics for both the general series and the day series are in Table \ref{PACDdescriptivestatistics}, indicating weekly periodicity, which supports the assumption $\nu = 7$.
\begin{table}
\centering
\caption{Descriptive Statistics }
\begin{tabular}{ccccccc}
\toprule
Day         & Minimum & Maximum & Mean   & Std. Dev. & Kurtosis & Skewness \\
\midrule
Full Series & 0.0900  & 43.7500 & 3.2043 & 3.8512    & 31.5103  & 4.1588   \\
Monday      & 0.3000  & 45.7500 & 4.0381 & 5.4249    & 32.9928  & 4.8292   \\
Tuesday     & 0.2700  & 20.5200 & 3.7044 & 3.6645    & 7.3127   & 1.9739   \\
Wednesday   & 0.4000  & 28.7300 & 3.6942 & 3.7656    & 15.9203  & 2.9831   \\
Thursday    & 0.3500  & 34.2500 & 3.4938 & 4.1828    & 26.7502  & 4.2428   \\
Friday      & 0.3200  & 24.0200 & 3.6212 & 3.5146    & 9.7172   & 2.1636   \\
Saturday    & 0.1100  & 22.5800 & 1.9197 & 2.8934    & 27.2164  & 4.5049   \\
Sunday      & 0.0900  & 11.6800 & 1.9590 & 1.8515    & 8.4375   & 1.9963   \\
\bottomrule
\end{tabular}
\label{PACDdescriptivestatistics}
\end{table}

\begin{figure}
    \centering
    \begin{subfigure}[b]{0.45\textwidth}
        \centering
        \includegraphics[width=\textwidth, height=3.2cm]{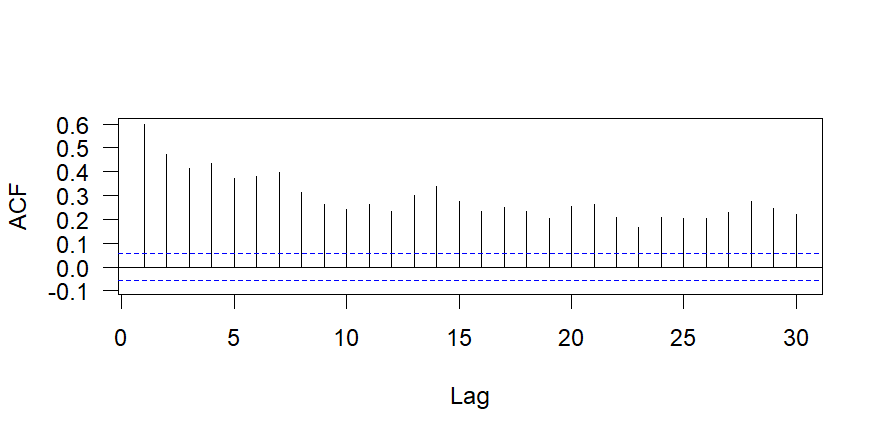}
        \caption{}
        \label{acfpacddata}
    \end{subfigure}
    \hfill
    \begin{subfigure}[b]{0.45\textwidth}
        \centering
        \includegraphics[width=\textwidth]{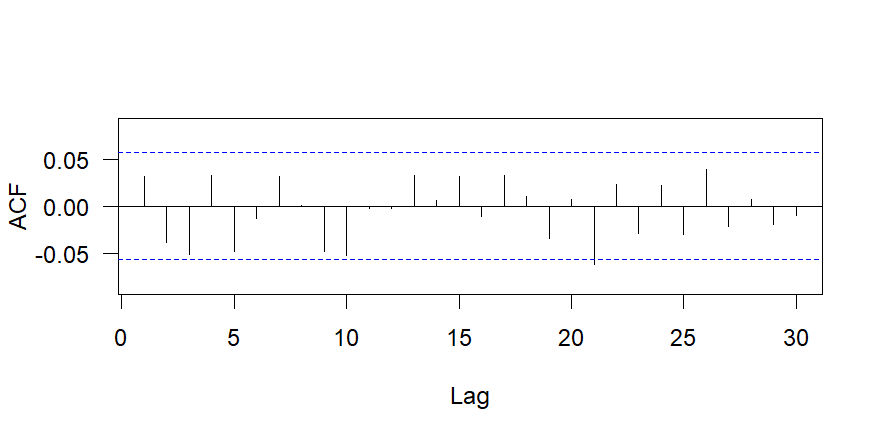}
        \caption{}
        \label{acfpacdres}
    \end{subfigure}
    \begin{subfigure}[b]{0.45\textwidth}
        \centering
        \includegraphics[width=\textwidth]{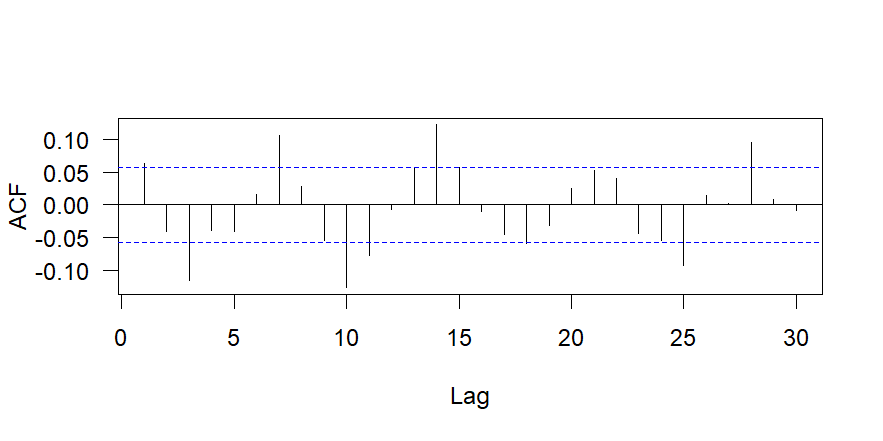}
        \caption{}
        \label{acffoupacdres}
    \end{subfigure}
    \hfill
    \begin{subfigure}[b]{0.45\textwidth}
        \centering
        \includegraphics[width=\textwidth]{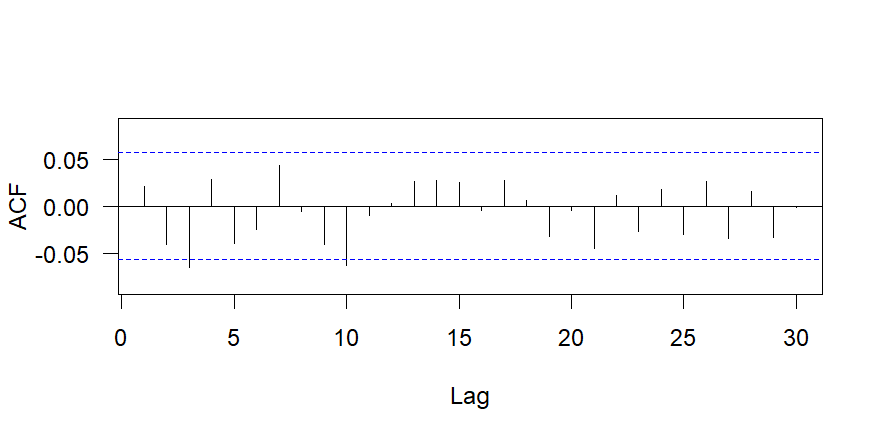}
        \caption{}
        \label{acfwavpacdres}
    \end{subfigure}
    \caption{(a) ACF plot of data. (b) ACF plot of $\text{PACD}_{7}(1,1)$ residuals. (c) ACF plot of Fourier-$\text{PACD}_{7}(1,1)$ residuals. (d) ACF plot of wavelet-$\text{PACD}_{7}(1,1)$ residuals.}
\end{figure}

The ACF plot in Figure \ref{acfpacddata} reveals strong serial autocorrelation in the data, which is further supported by Ljung-Box test results at lags 20 and 30, both yielding p-values less than $2.2 \times 10^{-16}$. 
From the 1204 observations, the first $171 \times 7 = 1197$ were used for estimation, and the final 7 observations were reserved for forecast comparisons. Both the initial values $u_{-1}$ and $\psi_{-1}$ were set to $y_{-1+7} = y_6$. The initial values for the nonlinear optimization used to compute the QMLE in \eqref{PACDQMLE} were obtained by minimizing the expression with the Genetic Algorithm.

Figure \ref{acfpacdres} presents the ACF plot of residuals from the $\text{PACD}_7(1,1)$ model. The Ljung-Box test yields p-values of 0.2659 and 0.2510 at lags 20 and 30, respectively, indicating no significant autocorrelation in the residuals.

Table \ref{foupacdanalysisres} shows the estimation results for the $ \text{Fourier-PACD}_7(1,1) $ model. 
\begin{table}
\caption{Estimation Results for Fourier-$\text{PACD}_7(1,1)$.}
\begin{tabular}{cccccc}
\toprule
  & \multicolumn{2}{c}{$\mathbf{f}_{\boldsymbol{\lambda}}$} &  & \multicolumn{2}{c}{$\mathbf{f}_{\boldsymbol{\gamma}}$}   \\
  \cmidrule{2-3} \cmidrule{5-6}
  & Estimate                    & Z-Score                   &  & Estimate                    & Z-Score                    \\
0 & \textbf{0.0994}                      & -                         &  & \textbf{0.3980}                      & -                          \\
1 & -0.0649                     & -1.13                     &  & -0.0784                     & -2.02                      \\
2 & 0.0471                      & 0.74                      &  & 0.0469                      & 0.72                       \\
3 & -0.0205                     & -0.23                     &  & -0.1018                     & -1.70                      \\
4 & -0.0030                     & -0.03                     &  & -0.0164                     & -0.33                      \\
5 & -0.0040                     & -0.04                     &  & 0.1167                      & 2.27                       \\
6 & 0.0535                      & 0.41                      &  & -0.1123                     & -1.84                      \\
  & \multicolumn{2}{c}{$\mathbf{f}_{\boldsymbol{\delta}}$}  &  & \multicolumn{2}{c}{$\mathbf{f}_{\boldsymbol{\sigma^2}}$} \\
   \cmidrule{2-3} \cmidrule{5-6}
0 & \textbf{0.6386}                      & -                         &  & \textbf{0.4301}                      & -                          \\
1 & -0.0736                     & -1.55                     &  & 0.0162                      & 0.19                       \\
2 & \textbf{0.2194*}                      & 2.87                      &  & 0.0951                      & 1.47                       \\
3 & \textbf{-0.2845*}                     & -3.61                     &  & 0.0279                      & 0.37                       \\
4 & -0.0887                     & -1.21                     &  & -0.0508                     & -0.69                      \\
5 & -0.1200                     & -1.54                     &  & 0.1013                      & 1.61                       \\
6 & -0.1312                     & -1.43                     &  & 0.0758                      & 0.89    \\      \bottomrule            
\end{tabular}
\footnotetext{The Fourier coefficients which are statistically significant $(|Z_{\mathbf{f},i}|>2.64)$ are indicated in *. The $\text{Fourier-PGARCH}_7(1,1)$ coefficients are written in bold.}
\label{foupacdanalysisres}
\end{table}
The estimated Fourier model is
\begin{equation}
\begin{split}
\hat{\lambda}_t &= 0.0994,\\
\hat{\gamma}_t &= 0.3980,\\
\hat{\delta}_t &= 0.6386+0.2194\sin\left(\frac{2 \pi t}{7}\right)-0.2845\cos\left(\frac{4 \pi t}{7}\right),\\
\text{and} \quad \hat{\sigma}_t^2 &=  0.4301, \, t=0,1,\dotsc,6.
\end{split}
\end{equation}

The ACF plot for its residuals is shown in Figure \ref{acffoupacdres}. The Ljung-Box test gives p-values of $7.47 \times 10^{-14}$ and $3.33 \times 10^{-16}$ at lags 20 and 30, respectively, revealing significant residual autocorrelation. This suggests that, although parsimonious with only 6 parameters, the Fourier-based model does not adequately capture the data structure.

Since the vectors $\boldsymbol{\lambda}, \boldsymbol{\gamma}, \boldsymbol{\delta}$, and $\boldsymbol{\sigma^2}$ have lengths of 7, which is not a power of two required for the computation of DWT, each vector was periodically extended to a length of 8, as outlined in Remark \ref{remark.per.ext}.
Among the Daubechies extremal phase and least asymmetric wavelets available in the \texttt{wavethresh} package in R, only the D(5) wavelet yielded a $\text{wavelet-PACD}_7(1,1)$ model that could capture the dependency structure based on the Ljung-Box test at the 5\% significance level. Table \ref{wavpacdanalysisres} summarizes the results of the D(5) DWT analysis. 
\begin{table}
\caption{Estimation Results for wavelet-$\text{PACD}_7(1,1)$.}
\begin{tabular}{cccccc}
\toprule
  & \multicolumn{2}{c}{$\mathbf{W}_{\boldsymbol{\lambda}}$} &  & \multicolumn{2}{c}{$\mathbf{W}_{\boldsymbol{\gamma}}$}   \\
   \cmidrule{2-3} \cmidrule{5-6}
  & Estimate                        & Z-Score               &  & Estimate                        & Z-Score                \\
0 & \textbf{0.2494}                 & -                     &  & \textbf{1.1032}                 & -                      \\
1 & 0.0143                          & 0.15                  &  & 0.0189                          & 0.21                   \\
2 & 0.1425                          & 0.98                  &  & 0.2124                          & 1.59                   \\
3 & -0.0963                         & -0.71                 &  & -0.0569                         & -0.73                  \\
4 & -0.0083                         & -0.22                 &  & 0.0302                          & 1.05                   \\
5 & -0.0471                         & -0.26                 &  & 0.3008                          & 2.25                   \\
6 & -0.0876                         & -0.36                 &  & 0.1113                          & 1.26                   \\
7 & -0.0361                         & -0.17                 &  & -0.0960                         & -1.10                  \\
  & \multicolumn{2}{c}{$\mathbf{W}_{\boldsymbol{\delta}}$}  &  & \multicolumn{2}{c}{$\mathbf{W}_{\boldsymbol{\sigma^2}}$} \\
   \cmidrule{2-3} \cmidrule{5-6}
0 & \textbf{1.6371}                 & -                     &  & \textbf{1.2679}                 & -                      \\
1 & -0.0640                         & -0.60                 &  & -0.2088                         & -1.71                  \\
2 & \textbf{0.8563*}                & 5.30                  &  & -0.0545                         & -0.32                  \\
3 & 0.0993                          & 1.01                  &  & -0.0657                         & -0.49                  \\
4 & 0.0129                          & 0.38                  &  & -0.0432                         & -1.03                  \\
5 & 0.0593                          & 0.29                  &  & 0.1189                          & 0.87                   \\
6 & 0.1500                          & 1.09                  &  & -0.2106                         & -1.07                  \\
7 & 0.1884                          & 1.53                  &  & -0.0872                         & -0.82 \\
\bottomrule
\end{tabular}
\footnotetext{The wavelet coefficients which are statistically significant $(|Z_{\mathbf{W},i}|>2.69)$ are indicated in *. The $\text{wavelet-PACD}_7(1,1)$ coefficients are written in bold.}
\label{wavpacdanalysisres}
\end{table}
The estimated DWT of $\boldsymbol{\lambda}, \boldsymbol{\gamma}, \boldsymbol{\delta}$ and $\boldsymbol{\sigma^2}$ are $\hat{\mathbf{W}}_{\boldsymbol{\lambda}} = (0.2494,0,\dotsc,0)^T, \hat{\mathbf{W}}_{\boldsymbol{\gamma}} = (1.1032,0,\dotsc,0)^T, \hat{\mathbf{W}}_{\boldsymbol{\delta}} = (1.6371,0,0.8563,0,\dotsc,0)^T$, and $\hat{\mathbf{W}}_{\boldsymbol{\sigma^2}} = (1.2679,0,\dotsc,0)^T$, respectively. Ljung-Box p-values for the residuals at lags 20 and 30 are 0.1900 and 0.2648, respectively, confirming no significant autocorrelation. Figure \ref{acfwavpacdres} depicts the corresponding ACF plot of residuals.

In terms of forecasting, the \( \text{wavelet-PACD}_7(1,1) \) model, consisting of only 5 parameters, outperformed the original model by achieving forecast accuracy \textbf{gains} of 11.47\% and 10.87\% in terms of RMSFE and MAFE, respectively. These findings underscore its parsimony and superior forecasting ability. This makes $\text{wavelet-PACD}_7(1,1)$, the most effective model among all those considered here.

\section{Conclusion}
Many financial time series display seasonal behavior in their characteristic features, such as conditional heteroscedasticity and conditional duration. In such cases, models like PGARCH and PACD are commonly used. However, these models often involve a substantial number of parameters, which can limit their efficiency. To address this, we propose Fourier and wavelet-based techniques to obtain more parsimonious versions of the PGARCH and PACD models. The effectiveness of these techniques is examined through simulation studies, while their practical relevance is illustrated using real-world data. The methods presented here can also be applied to other models with a large number of parameters.
\bmhead{Acknowledgements}
Rhea Davis gratefully acknowledges financial assistance from the University Grants Commission (UGC), India, through the Savitribai Jyotirao Phule Fellowship for Single Girl Child (SJSGC) scheme.
\section*{Declarations}
\begin{itemize}
\item \textbf{Conflict of interest} On behalf of all authors, the corresponding author states that there is no conflict of interest. 
\end{itemize}
\begin{appendices}
\section{}\label{secA1}
This section includes supporting materials related to the main text.\\
\hfill\\
\textbf{Estimation and Forecasting with PGARCH}\\
\hfill\\
  Let $y_0, y_1, \dotsc, y_{N\nu-1}$ be the data generated from \eqref{PGARCHeq}
  and $\boldsymbol{\theta} \in \Theta \subset ((0,\infty) \times [0,\infty)^2)^{\nu}$. Given initial values $y_{-1}$ and $\check{h}_{-1}$, let 
  \begin{equation*}
      \check{h}_{n\nu+k}(\boldsymbol{\theta}) = \omega_k+\alpha_ky_{n\nu+k-1}^2+\beta_k\check{h}_{n\nu+k-1}.
  \end{equation*}\
  Since the true distribution of the innovation sequence $\{\epsilon_t\}$ is not known, the quasi-maximum likelihood estimation method is used to estimate the parameter $\boldsymbol{\theta}$. For convenience, the form of the likelihood function is chosen to be Gaussian. The quasi-maximum likelihood estimate (QMLE) of $\boldsymbol{\theta}$, $\hat{\boldsymbol{\theta}}$, is (\citeauthor{aknouche2009quasi}, \citeyear{aknouche2009quasi}),
      \begin{equation}\label{PGARCHQMLE}
          \hat{\boldsymbol{\theta}} = \arg \min_{\boldsymbol{\theta} \in \boldsymbol{\Theta}} \frac{1}{N\nu}\sum_{n = 0}^{N-1} \sum_{k = 0}^{\nu-1} \left(\log(\check{h}_{n\nu+k})+\frac{y_{n\nu+k}^2}{\check{h}_{n\nu+k}}\right).
      \end{equation}
      \begin{equation*}
         \text{Let} \,\, A_{n\nu+k} = \begin{bmatrix}
\alpha_k\epsilon_{n\nu+k}^2 & \beta_k\epsilon_{n\nu+k}^2 \\
\alpha_k & \beta_k     
          \end{bmatrix}.
      \end{equation*}
      Define $A: = (A_{n\nu+k}, n,k \in \mathbb{Z})$. Let $M^2$ be the space of $2 \times 2$ real square matrices and $||\cdot||$ be a norm defined in this space. If $\sum\limits_{k = 0}^{\nu-1}E(\log^+||A_k||)<\infty$, where $\log^+ = \max(\log(w),0)$, then top Lyapunov exponent associated with A is defined by 
      \begin{equation*}
          \text{L}_{\nu}(A) = \inf_{n \in N^*}\frac{1}{n}E\{\log||A_0\dotsc A_{n\nu-1}||\}, 
      \end{equation*}
      where $N^*$ is the set of natural numbers. Also, $\text{L}_{\nu}(A_{\text{Tr}})$ denotes the top Lyapunov exponent when $A$ is determined under the true parameter $\boldsymbol{\theta}_{\text{Tr}}$. 
      Consider the following assumptions:\\ 
      a1. $\text{L}_{\nu}(A_{\text{Tr}})<0$ and for every $\boldsymbol{\theta} \in \boldsymbol{\Theta}$, $\beta_0 \times \dotsc \times \beta_{\nu-1}<1$.  \\
      a2. $\alpha_k \neq 0,$ for all $0\leq k\leq\nu-1$.\\
      a3. $\{\epsilon_t^2\}$ is non-degenerate. \\
      a4. $\boldsymbol{\theta}_{\text{Tr}}$ is in the interior of $\boldsymbol{\Theta}$.\\
      a5. $E(\epsilon_t^4)<\infty$.\\
      a6. The matrix $D = \sum \limits_{k=0}^{\nu-1}E_{\boldsymbol{\theta}_{\text{Tr}}}\left(\frac{1}{h_k^2(\boldsymbol{\theta}_{\text{Tr}})} \frac{\partial h_k(\boldsymbol{\theta}_{\text{Tr}})}{\partial \boldsymbol{\theta}} \frac{\partial h_k (\boldsymbol{\theta}_{\text{Tr}})}{\partial \boldsymbol{\theta}^T} \right)$ is finite and invertible.\\
\hfill\\
      Under assumptions a1-a3, $\hat{\boldsymbol{\theta}} \xrightarrow{a.s.}\boldsymbol{\theta}_{\text{Tr}}$ as $N \rightarrow \infty$, where $\xrightarrow{a.s.}$ denotes almost sure convergence. Also, the asymptotic normality of $\hat{\boldsymbol{\theta}}$ is stated in the following theorem (\citeauthor{aknouche2009quasi}, \citeyear{aknouche2009quasi}).
      \begin{theorem}\label{PGARCHasynormthm}
      Under assumptions a1-a6, 
      \begin{equation}\label{PGARCHasynorm}
                \sqrt{N}(\hat{\boldsymbol{\theta}}-\boldsymbol{\theta}_{\text{Tr}}) \xrightarrow{d} \mathcal{N}(\mathbf{0}, \Sigma) \quad \text{as} \quad N \rightarrow \infty,    
      \end{equation}
      where $\Sigma = (E(\epsilon_t^4)-1)D^{-1}$ is a block diagonal matrix
      and $\xrightarrow{d}$ denotes the convergence in distribution. 
      \end{theorem}
      Let $\Sigma = [s_{v,w}]_{0\leq v\leq3\nu-1, 0\leq w\leq3\nu-1}$.
  \begin{corollary}\label{PGARCHestasynormthm}
  From \eqref{PGARCHasynorm}, we have, 
 \begin{equation}\label{omega.asynorm}
              (\text{i})\quad \sqrt{N}(\hat{\boldsymbol{\omega}}-\boldsymbol{\omega}) \xrightarrow{d} \mathcal{N}(\mathbf{0}, \Sigma_{\boldsymbol{\omega}}) \quad \text{as} \quad N \rightarrow \infty,
          \end{equation}
          where 
          \begin{equation}
            \Sigma_{\boldsymbol{\omega}} = \begin{bmatrix}
            s_{0,0} & s_{0,3} & s_{0,6} & \dotsc & s_{0,3\nu-3}\\
             s_{3,0} & s_{3,3} & s_{3,6} & \dotsc & s_{3,3\nu-3}\\
             \vdots & \vdots & \vdots & \ddots & \vdots \\
            s_{3\nu-3,0} & s_{3\nu-3,3} & s_{3\nu-3,6} & \dotsc & s_{3\nu-3,3\nu-3}
            \end{bmatrix}.
          \end{equation}
          \begin{equation}
           (\text{ii})\quad    \sqrt{N}(\hat{\boldsymbol{\alpha}}-\boldsymbol{\alpha}) \xrightarrow{d} \mathcal{N}(\mathbf{0}, \Sigma_{\boldsymbol{\alpha}}) \quad \text{as} \quad N \rightarrow \infty,
          \end{equation}
          where 
          \begin{equation}
            \Sigma_{\boldsymbol{\alpha}} = \begin{bmatrix}
            s_{1,1} & s_{1,4} & s_{1,7} & \dotsc & s_{1,3\nu-2}\\
             s_{4,1} & s_{4,4} & s_{4,7} & \dotsc & s_{4,3\nu-2}\\
             \vdots & \vdots & \vdots & \ddots & \vdots \\
            s_{3\nu-2,1} & s_{3\nu-2,4} & s_{3\nu-2,7} & \dotsc & s_{3\nu-2,3\nu-2}
            \end{bmatrix}.
          \end{equation}
          \begin{equation}
            (\text{iii})\quad   \sqrt{N}(\hat{\boldsymbol{\beta}}-\boldsymbol{\beta}) \xrightarrow{d} \mathcal{N}(\mathbf{0}, \Sigma_{\boldsymbol{\beta}}) \quad \text{as} \quad N \rightarrow \infty,
          \end{equation}
          where 
          \begin{equation}
            \Sigma_{\boldsymbol{\beta}} = \begin{bmatrix}
            s_{2,2} & s_{2,5} & s_{2,8} & \dotsc & s_{2,3\nu-1}\\
             s_{5,2} & s_{5,5} & s_{5,8} & \dotsc & s_{5,3\nu-1}\\
             \vdots & \vdots & \vdots & \ddots & \vdots \\
            s_{3\nu-1,2} & s_{3\nu-1,5} & s_{3\nu-1,8} & \dotsc & s_{3\nu-1,3\nu-1}
            \end{bmatrix}.
          \end{equation}
  \end{corollary}
  The asymptotic variance-covariance matrix $\Sigma$ in \eqref{PGARCHasynorm} is estimated by plugging in the estimates of $D$ and $E(\epsilon_t^4)$. $D$ is consistently estimated using 
  \begin{equation}
      \hat{D} = \frac{1}{N}\sum_{n=0}^{N-1} \sum_{k=0}^{\nu-1}\frac{1}{h_{n\nu+k}^2(\hat{\boldsymbol{\theta}})}\frac{\partial h_{n\nu+k}(\hat{\boldsymbol{\theta}})}{\partial \boldsymbol{\theta}}\frac{\partial h_{n\nu+k}(\hat{\boldsymbol{\theta}})}{\partial \boldsymbol{\theta}^T}.
  \end{equation}
  Additionally, $E(\epsilon_t^4)$ is estimated by its moment estimator, viz., $\frac{1}{N\nu}\sum\limits_{t=0}^{N\nu-1}r_{\hat{\boldsymbol{\theta}},t}^4$, where $r_{\hat{\boldsymbol{\theta}},t} = y_t/\sqrt{\hat{h}_t}$, $\hat{h}_t = \hat{\omega}_t+\hat{\alpha}_ty_{t-1}^2+\hat{\beta}_th_{t-1}$. Note that $\{r_{\hat{\boldsymbol{\theta}},t}\}$ are the residuals of the PGARCH model \eqref{PGARCHeq}.

  The 1-step ahead forecast of $\text{PGARCH}_{\nu}(1,1)$ is obtained by
          \begin{align}
              h_{N\nu-1}(1) &= \omega_{N\nu}+\alpha_{N\nu}y_{N\nu-1}^2+\beta_{N\nu}h_{N\nu-1} \nonumber\\
              &= \omega_{0}+\alpha_{0}y_{N\nu-1}^2+\beta_{0}h_{N\nu-1}, 
          \end{align}
          and the $\ell$-step ($\ell>1$) ahead forecast is
          \begin{equation}
              h_{N\nu-1}(\ell) = \omega_{N\nu-1+\ell}+(\alpha_{N\nu-1+\ell}+\beta_{N\nu-1+\ell})h_{N\nu-1}(\ell-1).
          \end{equation}
          The following section describes the estimation and forecasting with the PACD model.\\
          \hfill\\
\textbf{Estimation and Forecasting with PACD}\\
\hfill\\
Consider the data $u_0, \dotsc, u_{N\nu-1}$ obtained from \eqref{PACD eq}
          and let $\boldsymbol{\eta} = (\eta_0, \dotsc, \eta_{3\nu-1})^T$ \\  
          $= (\lambda_0, \gamma_0, \delta_0,\dotsc, \lambda_{\nu-1}, \gamma_{\nu-1}, \delta_{\nu-1})^T \in \text{\textbf{H}} \subset ((0,\infty) \times [0,\infty)^2)^{\nu}$. If $u_{-1}$ and $\check{\psi}_{-1}$ are known, let 
          \begin{equation*}
              \check{\psi}_{n\nu+k}(\boldsymbol{\eta}) = \lambda_k+\gamma_ku_{n\nu+k-1}^2+\delta_k\check{\psi}_{n\nu+k-1}.
          \end{equation*}
Here, we do not assume a particular distribution for $\{\xi_t\}$. For positive-valued time series, it is often found that QMLE based on the exponential distribution is a reliable estimator for $\boldsymbol{\eta}$. The exponential quasi-maximum likelihood (EQMLE) $\hat{\boldsymbol{\eta}}$ of $\boldsymbol{\eta}$ is (\citeauthor{aknouche2022periodic}, \citeyear{aknouche2022periodic})
          \begin{equation}\label{PACDQMLE}
              \hat{\boldsymbol{\eta}} = \arg \min_{\boldsymbol{\eta} \in \textbf{H}} \frac{1}{N\nu} \sum _{n = 0}^{N-1} \sum _{k = 0}^{\nu-1} \left(\log(\check{\psi}_{n\nu+k})+\frac{u_{n\nu+k}}{\check{\psi}_{n\nu+k}}\right).
      \end{equation}
  \begin{equation*}
         \text{Let} \,\, B_{n\nu+k} = \begin{bmatrix}
\gamma_k\xi_{n\nu+k} & \delta_k\xi_{n\nu+k} \\
\gamma_k & \delta_k     
          \end{bmatrix}.
      \end{equation*}
      Define $B: = (B_{n\nu+k}, n,k \in \mathbb{Z})$. The properties of $ \hat{\boldsymbol{\eta}}$ are established under the following assumptions.\\
      b1. $\sum\limits_{k = 0}^{\nu-1}E(\log^+||B_k| |)<\infty$, $\text{L}_{\nu}(B_{\text{Tr}})<0$, where $B$ computed using the true parameter $\boldsymbol{\eta}_{\text{Tr}}$ is denoted by $B_{\text{Tr}}$. Also, $\delta_0 \times \dotsc \times \delta_{\nu-1}<1,$ for every $\boldsymbol{\eta} \in \text{\textbf{H}}$.\\
      b2. $\boldsymbol{\eta} \in \textbf{H}$ and $\textbf{H}$ is compact.\\
      b3. $\gamma_k \neq 0,$ for $0\leq k \leq \nu-1.$\\
      b4. $\{\xi_k\}$ is non-degenerate $E(\xi_k) = 1$, for all $0 \leq k \leq \nu-1$. \\
      b5. $\boldsymbol{\eta}_{Tr}$ belongs to the interior of \textbf{H}.\\
   b6. $Var(\xi_k) = \sigma_k^2 \in (0,\infty)$, for all $0 \leq k \leq \nu-1$, the matrices $G = \sum\limits_{k=0}^{\nu-1}E\left(\frac{1}{\psi_k^2(\boldsymbol{\eta}_{\text{Tr}})} \frac{\partial \psi_k(\boldsymbol{\eta}_{\text{Tr}})}{\partial \boldsymbol{\eta}}\frac{\partial \psi_k(\boldsymbol{\eta}_{\text{Tr}})}{\partial \boldsymbol{\eta}^T}\right)$ and $K = \sum\limits_{k=0}^{\nu-1}\sigma_k^2 E\left(\frac{1}{\psi_k^2(\boldsymbol{\eta}_{\text{Tr}})} \frac{\partial \psi_k(\boldsymbol{\eta}_{\text{Tr}})}{\partial \boldsymbol{\eta}}\frac{\partial \psi_k(\boldsymbol{\eta}_{\text{Tr}})}{\partial \boldsymbol{\eta}^T}\right)$ are finite and $G$ is nonsingular.\\
    \hfill\\
  Under assumptions b1-b4, $\hat{\boldsymbol{\eta}} \xrightarrow{a.s.}\boldsymbol{\eta}_{\text{Tr}}$ as $N \rightarrow \infty$. Moreover, $\hat{\boldsymbol{\eta}}$ has asymptotic normality as stated in the subsequent theorem (\citeauthor{aknouche2022periodic}, \citeyear{aknouche2022periodic}).
  \begin{theorem}
      Under assumptions b1-b6, 
      \begin{equation}\label{PACDasynorm}
      \sqrt{N}(\hat{\boldsymbol{\eta}}-\boldsymbol{\eta}_{\text{Tr}}) \xrightarrow{d} \mathcal{N}(\mathbf{0}, \Xi) \quad \text{as} \quad N \rightarrow \infty,    
      \end{equation}
  where $\Xi = G^{-1}KG^{-1}$ is a block diagonal matrix. 
  \end{theorem}
  Let $\Xi = [e_{a,b}]_{0\leq a\leq 3\nu-1, 0\leq b\leq 3\nu-1}$.
  \begin{corollary}
      From \eqref{PACDasynorm}, we have, 
          \begin{equation}
             (\text{i})\quad  \sqrt{N}(\hat{\boldsymbol{\lambda}}-\boldsymbol{\lambda}) \xrightarrow{d} \mathcal{N}(\mathbf{0}, \Xi_{\boldsymbol{\lambda}}) \quad \text{as} \quad N \rightarrow \infty,
          \end{equation}
          where 
          \begin{equation}
            \Xi_{\boldsymbol{\lambda}} = \begin{bmatrix}
            e_{0,0} & e_{0,3} & e_{0,6} & \dotsc & e_{0,3\nu-3}\\
             e_{3,0} & e_{3,3} & e_{3,6} & \dotsc & e_{3,3\nu-3}\\
             \vdots & \vdots & \vdots & \ddots & \vdots \\
            e_{3\nu-3,0} & e_{3\nu-3,3} & e_{3\nu-3,6} & \dotsc & e_{3\nu-3,3\nu-3}
            \end{bmatrix}.
          \end{equation}
          \begin{equation}
              (\text{ii})\quad  \sqrt{N}(\hat{\boldsymbol{\gamma}}-\boldsymbol{\gamma}) \xrightarrow{d} \mathcal{N}(\mathbf{0}, \Xi_{\boldsymbol{\gamma}}) \quad \text{as} \quad N \rightarrow \infty,
          \end{equation}
          where 
          \begin{equation}
            \Xi_{\boldsymbol{\gamma}} = \begin{bmatrix}
            e_{1,1} & e_{1,4} & e_{1,7} & \dotsc & e_{1,3\nu-2}\\
             e_{4,1} & e_{4,4} & e_{4,7} & \dotsc & e_{4,3\nu-2}\\
             \vdots & \vdots & \vdots & \ddots & \vdots \\
            e_{3\nu-2,1} & e_{3\nu-2,4} & e_{3\nu-2,7} & \dotsc & e_{3\nu-2,3\nu-2}
            \end{bmatrix}.
          \end{equation}
          \begin{equation}
           (\text{iii})\quad    \sqrt{N}(\hat{\boldsymbol{\delta}}-\boldsymbol{\delta}) \xrightarrow{d} \mathcal{N}(\mathbf{0}, \Xi_{\boldsymbol{\delta}}) \quad \text{as} \quad N \rightarrow \infty,
          \end{equation}
          where 
          \begin{equation}
            \Xi_{\boldsymbol{\delta}} = \begin{bmatrix}
            e_{2,2} & e_{2,5} & e_{2,8} & \dotsc & e_{2,3\nu-1}\\
             e_{5,2} & e_{5,5} & e_{5,8} & \dotsc & e_{5,3\nu-1}\\
             \vdots & \vdots & \vdots & \ddots & \vdots \\
            e_{3\nu-1,2} & e_{3\nu-1,5} & e_{3\nu-1,8} & \dotsc & e_{3\nu-1,3\nu-1}
            \end{bmatrix}.
          \end{equation}
  \end{corollary}
  The innovation variance $\sigma_k^2, k = 0, 1, \dotsc, \nu-1$ is estimated via (\citeauthor{aknouche2022periodic}, \citeyear{aknouche2022periodic})
      \begin{equation}
          \hat{\sigma}_k^2 = \frac{1}{N}\sum_{n=0}^{N-1} \left(\frac{u_{n\nu+k}-\hat{\psi}_{n\nu+k}}{\hat{\psi}_{n\nu+k}}\right)^2, \, \, \,  \text{for all} \,\,\, 0 \leq k \leq \nu-1,
      \end{equation}
      where $\hat{\psi}_{n\nu+k} = \hat{\lambda}_k+\hat{\gamma}_ku_{n\nu+k-1}+\hat{\delta}_k\hat{\psi}_{n\nu+k-1}$. Under assumptions b1-b4, $\hat{\boldsymbol{\sigma}}^2 \xrightarrow{a.s.}\boldsymbol{\sigma}_{\text{Tr}}^{2}$ as $N \rightarrow \infty$. Moreover, $\hat{\boldsymbol{\sigma}}^2$ has asymptotic normality as stated in the theorem presented below (\citeauthor{aknouche2022periodic}, \citeyear{aknouche2022periodic}).
      \begin{theorem}
          Under assumptions b1-b4 and if $E(\xi_k^4) < \infty$ for all $0 \leq k \leq \nu-1$,
          \begin{equation}
              \sqrt{N}(\hat{\boldsymbol{\sigma}}^2- \boldsymbol{\sigma}_{\text{Tr}}^2) \xrightarrow{d} \mathcal{N}(\mathbf{0},\Lambda) \,\, \text{as}\,\, N \rightarrow \infty,
          \end{equation}
          where $\Lambda = \text{diag}(\Lambda_0, \dotsc, \Lambda_{\nu-1})$ and $\Lambda_{k} = E((\xi_k-1)^2-\sigma_k^2)^2.$
      \end{theorem}
      A consistent estimator of $\Lambda_k, 0 \leq k \leq \nu-1$, is (\citeauthor{aknouche2022periodic}, \citeyear{aknouche2022periodic})
      \begin{equation}
          \hat{\Lambda}_k = \frac{1}{N}\sum_{n=0}^{N-1} ((\hat{\xi}_{n\nu+k}-1)^2-\hat{\sigma}_k^2)^2, \quad 0 \leq k \leq \nu-1,
      \end{equation}
      where $\hat{\xi}_{n\nu+k} = \frac{u_{n\nu+k}}{\hat{\psi}_{n\nu+k}}$ are the residuals.
   $G$ and $K$ of \eqref{PACDasynorm} are consistently estimated using (\citeauthor{aknouche2022periodic}, \citeyear{aknouche2022periodic})
  \begin{equation}
      \hat{G} = \frac{1}{N}\sum_{n=0}^{N-1}\sum_{k=0}^{\nu-1} \frac{1}{\psi_{n\nu+k}^2(\hat{\boldsymbol{\eta}})}\frac{\partial \psi_{n\nu+k}(\hat{\boldsymbol{\eta}})}{\partial \boldsymbol{\eta}}\frac{\partial \psi_{n\nu+k}(\hat{\boldsymbol{\eta}})}{\partial \boldsymbol{\eta}^T},
  \end{equation}
  and 
  \begin{equation}
      \hat{K} = \frac{1}{N}\sum_{n=0}^{N-1}\sum_{k=0}^{\nu-1} \frac{\hat{\sigma}_k^2}{\psi_{n\nu+k}^2(\hat{\boldsymbol{\eta}})}\frac{\partial \psi_{n\nu+k}(\hat{\boldsymbol{\eta}})}{\partial \boldsymbol{\eta}}\frac{\partial \psi_{n\nu+k}(\hat{\boldsymbol{\eta}})}{\partial \boldsymbol{\eta}^T},
  \end{equation}
  respectively. Thus, the estimator of $\Xi$, $\hat{\Xi}$, is obtained using $\hat{\Xi} = (\hat{G})^{-1}\hat{K}(\hat{G})^{-1}.$ 
      
  The 1-step ahead forecast of $\text{PACD}_{\nu}(1,1)$ is computed by
          \begin{align}
              \psi_{N\nu-1}(1) &= \lambda_{N\nu}+\gamma_{N\nu}u_{N\nu-1}+\delta_{N\nu}\psi_{N\nu-1} \nonumber\\
              &= \lambda_{0}+\gamma_{0}u_{N\nu-1}+\delta_{0}\psi_{N\nu-1},
          \end{align}
          and the $\ell$-step ($\ell>1$) ahead forecast is
          \begin{equation}
              \psi_{N\nu-1}(\ell) = \lambda_{N\nu-1+\ell}+(\gamma_{N\nu-1+\ell}+\delta_{N\nu-1+\ell})\psi_{N\nu-1}(\ell-1).
          \end{equation}
          The following theorem demonstrates the asymptotic normality of a transformed vector (\citeauthor{brockwell1991time}, \citeyear{brockwell1991time}).
\begin{theoremA}
    \label{A.broc.asymnormaltheorem}
If $\mathbf{X}_n$ is a $k \times 1$ vector such that $\mathbf{X}_n$ is AN$(\boldsymbol{\mu_n},\Sigma_n)$
and $\mathcal{B}$ is any non-zero $m \times k$ matrix such that the matrix $\mathcal{B}\Sigma_n \mathcal{B}^{T}$ have no zero diagonal elements, then 
$\mathcal{B}\mathbf{X}_n$  is AN$(\mathcal{B}\boldsymbol{\mu}_n,\mathcal{B}\Sigma_n \mathcal{B}^{T}).$
\end{theoremA}
\section{}
The Haar transform matrix of order 8 is (rounded off to two decimal places)
\begin{equation}\label{Haarmat}
    \begin{bmatrix}
        0.35 & 0.35 & 0.35 & 0.35 & 0.35 & 0.35 & 0.35 & 0.35 \\ 
  0.35 & 0.35 & 0.35 & 0.35 & -0.35 & -0.35 & -0.35 & -0.35 \\ 
   0.50 & 0.50 & -0.50 & -0.50 & 0.00 & 0.00 & 0.00 & 0.00 \\ 
  0.00 & 0.00 & 0.00 & 0.00 & 0.50 & 0.50 & -0.50 & -0.50 \\ 
   0.71 & -0.71 & 0.00 & 0.00 & 0.00 & 0.00 & 0.00 & 0.00 \\ 
 0.00 & 0.00 & 0.71 & -0.71 & 0.00 & 0.00 & 0.00 & 0.00 \\ 
  0.00 & 0.00 & 0.00 & 0.00 & 0.71 & -0.71 & 0.00 & 0.00 \\ 
  0.00 & 0.00 & 0.00 & 0.00 & 0.00 & 0.00 & 0.71 & -0.71 \\ 
    \end{bmatrix}.
\end{equation}
The D(5) transform matrix of order 8 is (rounded off to two decimal places)
\begin{equation}\label{D5mat}
    \begin{bmatrix}
0.35 & 0.35 & 0.35 & 0.35 & 0.35 & 0.35 & 0.35 & 0.35 \\ 
-0.45 & -0.45 & -0.29 & 0.11 & 0.45 & 0.45 & 0.29 & -0.11 \\ 
-0.52 & 0.23 & 0.63 & 0.16 & -0.11 & 0.08 & 0.00 & -0.48 \\ 
 -0.11 & 0.08 & 0.00 & -0.48 & -0.52 & 0.23 & 0.63 & 0.16 \\ 
  0.61 & -0.15 & -0.01 & -0.08 & -0.03 & 0.24 & 0.14 & -0.72 \\ 
 0.14 & -0.72 & 0.61 & -0.15 & -0.01 & -0.08 & -0.03 & 0.24 \\ 
 -0.03 & 0.24 & 0.14 & -0.72 & 0.61 & -0.15 & -0.01 & -0.08 \\ 
-0.01 & -0.08 & -0.03 & 0.24 & 0.14 & -0.72 & 0.61 & -0.15 
    \end{bmatrix}.
    \end{equation}
The D(8) transform matrix of order 8 is (rounded off to two decimal places)
\begin{equation}\label{D8mat}
    \begin{bmatrix}
    0.35 & 0.35 & 0.35 & 0.35 & 0.35 & 0.35 & 0.35 & 0.35 \\ 
  0.33 & -0.02 & -0.36 & -0.51 & -0.33 & 0.02 & 0.36 & 0.51 \\ 
  -0.49 & -0.62 & 0.00 & 0.28 & 0.02 & 0.10 & 0.47 & 0.25 \\ 
  0.02 & 0.10 & 0.47 & 0.25 & -0.49 & -0.62 & 0.00 & 0.28 \\ 
0.27 & -0.04 & 0.13 & 0.00 & -0.28 & 0.02 & 0.59 & -0.69 \\ 
  0.59 & -0.69 & 0.27 & -0.04 & 0.13 & 0.00 & -0.28 & 0.02 \\ 
 -0.28 & 0.02 & 0.59 & -0.69 & 0.27 & -0.04 & 0.13 & 0.00 \\ 
  0.13 & 0.00 & -0.28 & 0.02 & 0.59 & -0.69 & 0.27 & -0.04 
    \end{bmatrix}.
\end{equation}
The LA(5) transform matrix of order 8 is (rounded off to two decimal places)
\begin{equation}\label{LA5mat}
    \begin{bmatrix}
 0.35 & 0.35 & 0.35 & 0.35 & 0.35 & 0.35 & 0.35 & 0.35 \\ 
-0.36 & -0.49 & -0.36 & -0.07 & 0.36 & 0.49 & 0.36 & 0.07 \\ 
 0.01 & 0.18 & -0.01 & 0.11 & 0.52 & 0.28 & -0.52 & -0.58 \\ 
 0.52 & 0.28 & -0.52 & -0.58 & 0.01 & 0.18 & -0.01 & 0.11 \\ 
0.05 & -0.01 & -0.18 & -0.02 & 0.63 & -0.72 & 0.20 & 0.04 \\ 
 0.20 & 0.04 & 0.05 & -0.01 & -0.18 & -0.02 & 0.63 & -0.72 \\ 
0.63 & -0.72 & 0.20 & 0.04 & 0.05 & -0.01 & -0.18 & -0.02 \\ 
 -0.18 & -0.02 & 0.63 & -0.72 & 0.20 & 0.04 & 0.05 & -0.01 
    \end{bmatrix}.
\end{equation}

\FloatBarrier
\begin{table}[htbp]
\caption{Performance of $\text{wavelet-PGARCH}_7(1,1)$ with different wavelets.}
\begin{tabular}{cccc}
\toprule
Model                 & Number of parameters & \multicolumn{2}{c}{Forecast Error Measures} \\

                      &                      & RMSFE                & MAFE                 \\
                      \midrule
\textbf{D(1) or Haar} & \textbf{3}           & \textbf{8.6593}      & \textbf{8.2990}      \\
\textbf{D(2)}         & \textbf{3}           & \textbf{8.6593}      & \textbf{8.2990}      \\
D(3)                  & 5                    & 12.3065              & 11.5738              \\
D(4)                  & 4                    & 9.5155               & 9.0517               \\
D(5)                  & 4                    & 9.9761               & 9.2452               \\
D(6)                  & 4                    & 8.5231               & 7.8671               \\
\textbf{D(7)}         & \textbf{3}           & \textbf{8.6593}      & \textbf{8.2990}      \\
D(8)                  & 4                    & 11.0470              & 10.2732              \\
D(9)                  & 5                    & 11.1047              & 10.4870              \\
D(10)                 & 5                    & 14.7204              & 13.9677              \\
LA(4)                 & 4                    & 9.3760               & 8.9851               \\
\textbf{LA(5)}        & \textbf{3}           & \textbf{8.6593}      & \textbf{8.2990}      \\
LA(6)                 & 4                    & 10.9480              & 10.1498              \\
LA(7)                 & 4                    & 10.5023              & 9.8365               \\
\textbf{LA(8)}        & \textbf{3}           & \textbf{8.6593}      & \textbf{8.2990}      \\
\textbf{LA(9)}        & \textbf{3}           & \textbf{8.6593}      & \textbf{8.2990}      \\
LA(10)                & 5                    & 10.4409              & 9.6280         \\
\bottomrule
\end{tabular}
\label{PGARCHdiffwavanalysis}
\end{table}
\FloatBarrier
\end{appendices}
\bibliography{references}
\end{document}